\documentclass[preprint,review,12pt,authoryear]{elsarticle}

\usepackage{amssymb}
\usepackage[dvipsnames]{xcolor}
\usepackage{soul} 
\usepackage{graphicx}
\usepackage{balance}
\usepackage{amsmath}
\usepackage[hyphens]{url}
\usepackage{setspace}
\usepackage[font=footnotesize,labelfont=bf]{caption}
\usepackage[font=footnotesize,labelfont=bf]{subcaption}
\def\BibTeX{{\rm B\kern-.05em{\sc i\kern-.025em b}\kern-.08em
    T\kern-.1667em\lower.7ex\hbox{E}\kern-.125emX}}

\journal{Expert Systems with Applications}

\begin{document}

\begin{frontmatter}



\title{A Compositional Model of Multi-faceted Trust for Personalized Item Recommendation}


\author{Liliana Ardissono\footnote{Corresponding author.\\
E-mail addresses: liliana.ardissono@unito.it (L. Ardissono), noemi.mauro@unito.it (N. Mauro).
} and Noemi Mauro}

\address{Computer Science Department, University of Torino \\ Corso Svizzera, 185, I-10149, Torino,  Italy}

\begin{abstract}
Trust-based recommender systems improve rating prediction with respect to Collaborative Filtering by leveraging the additional information provided by a trust network among users to deal with the cold start problem. 
However, they are challenged by recent studies according to which people generally perceive the usage of data about social relations as a violation of their own privacy. 
In order to address this issue, we extend trust-based recommender systems with additional evidence about trust, based on public anonymous information, and we make them configurable with respect to the data that can be used in the given application domain: 
\begin{enumerate}
    \item We propose the Multi-faceted Trust Model (MTM) to define trust among users in a compositional way, possibly including or excluding the types of information it contains. MTM flexibly integrates social links with public anonymous feedback received by user profiles and user contributions in social networks. 
    \item We propose LOCABAL+, based on MTM, which extends the LOCABAL trust-based recommender system with multi-faceted trust and trust-based social regularization.
\end{enumerate}
Experiments carried out on two public datasets of item reviews show that, with a minor loss of user coverage, LOCABAL+ outperforms state-of-the art trust-based recommender systems and Collaborative Filtering in accuracy, ranking of items and error minimization both when it uses complete information about trust and when it ignores social relations. 
The combination of MTM with LOCABAL+ thus represents a promising alternative to state-of-the-art trust-based recommender systems.
\end{abstract}


%

\begin{highlights}
\item 
We propose a compositional recommender system based on multi-faceted trust.
\item 
The trust model is based on social links and global feedback about users.
\item The recommender can work with or without using information about social relations.
\item 
We validate our recommendation model on two public datasets of item reviews.
\item
Our recommender outperforms state-of-the-art trust-based recommender systems.
\end{highlights}

\fontsize{10pt}{8pt}\selectfont\bfseries Published in Expert Systems with Applications, Elsevier. \\ DOI: https://doi.org/10.1016/j.eswa.2019.112880.
	Link to the page of the paper on Elsevier web site: \\
	http://www.sciencedirect.com/science/article/pii/S0957417419305901 \\
	This work is licensed under the Creative Commons Attribution-NonCommercial-NoDerivatives 4.0 International 
	License. To view a copy of this license, visit \\ http://creativecommons.org/licenses/by-nc-nd/4.0/ or send a letter to 
	Creative Commons, PO Box 1866, Mountain View, CA 94042, USA.\\


\end{frontmatter}


\noindent
Declarations of interest: none.
\begin{flushleft}
\fontsize{10pt}{8pt}\selectfont\bfseries Published in Expert Systems with Applications, Elsevier. \\ DOI: https://doi.org/10.1016/j.eswa.2019.112880.
Link to the page of the paper on Elsevier web site: \\
http://www.sciencedirect.com/science/article/pii/S0957417419305901 \\
This work is licensed under the Creative Commons Attribution-NonCommercial-NoDerivatives 4.0 International 
License. To view a copy of this license, visit \\ http://creativecommons.org/licenses/by-nc-nd/4.0/ or send a letter to 
Creative Commons, PO Box 1866, Mountain View, CA 94042, USA.\\
\end{flushleft}

\section{Introduction}
Trust-based recommender systems improve rating prediction with respect to Collaborative Filtering \citep{Desrosiers-Karypis:11} by combining rating similarity with the additional information provided by a trust network among users to deal with the cold start problem. 
Most of these systems predict the rating scores that a person would attribute to items by relying on the observed preferences of the users who are linked to her/him by social relations, directly or through a short path of links, as in SocialMF \citep{Jamali-Ester:10}.
Moreover, having observed that, in the physical world, people are likely to seek advice from both local friends and highly reputable users, some systems also take global reputation into account to improve recommendation performance. For instance, LOCABAL \citep{Tang-etal:13} computes users' reputation as a function of their importance in the social network and exploits this data to weight the impact of ratings in Matrix Factorization. 

Despite the good recommendation results achieved by trust-based recommender systems, recent studies show that they are hardly accepted by people, who are concerned about the storage of personal information and the access to social relations \citep{Burbach-etal:18}. It is thus vital to define trust models that can use data which is not perceived as personal.

For this purpose, we propose a compositional trust model and recommender system which rely on complementary information sources to obtain a twofold objective: (i) collecting rich evidence about user trust to improve Top-N recommendation, and (ii) adapting to possible restrictions on the information that can be used in the application domain of interest.

Indeed, various signs of trust can be used to compute users' global reputation without relying on sensitive information.
For instance, social networks such as \cite{Booking}, \cite{Expedia} and \cite{Yelp} publish anonymous feedback about users (expressed as endorsements to their profiles) and about their contributions (e.g., helpfulness of reviews) that can be used to assess reputation by ignoring the identity of the people who provided it. It is thus interesting to define a model that supports the interpretation of these types of feedback as an overall trustworthiness measure.

In this article we present the Multi-faceted Trust Model (MTM) as a framework to fuse {\em local trust between users} (inferred from direct social relations) with the following sources of information:
\begin{itemize}
    \item
    The {\em quality of individual reviews} derived from the explicit feedback they receive from the social network.
		\item 
    {\em Multi-dimensional user reputation} derived from the analysis and integration of different types of endorsements that users can receive with the quality of the reviews they author. 
\end{itemize}
MTM makes it possible to separately include or exclude the components of trust to assess their relative impact on recommendation. This supports the evaluation of performance, e.g., when different types of anonymous feedback are considered, and when social relations are ignored. 

We integrate MTM into a novel trust-based recommendation algorithm, denoted as LOCABAL+, which combines local trust and multi-dimensional global reputation in preference estimation. LOCABAL+ extends the LOCABAL recommender system, from which we take inspiration, as follows: 
\begin{itemize} 
\item It tunes Matrix Factorization by exploiting multi-faceted trust, which takes multiple aspects of user behavior into account, instead of only relying on social links. 
\item It regularizes social relations by means of rating similarity and multi-dimensional global reputation to exploit both properties in the selection of the like-minded users for rating prediction. 
\item 
It can be configured to use a subset of the facets of trust.
\end{itemize}
Experimental results show that LOCABAL+ achieves the best accuracy, error minimization and ranking results when it uses both global trust feedback and social relations. However, it also outperforms state-of-the-art recommender systems based on Matrix Factorization and on K-Nearest Neighbors when it ignores social relations. We thus conclude that multi-faceted trust enhances recommendation performance in trust-based recommenders and it makes them more flexible with respect to the types of information that can be used in a specific application domain. 

This work extends the preliminary multi-faceted trust model presented in \citep{Mauro-etal:19} as follows:
firstly, we integrate the facets of trust in Matrix Factorization, instead of using a K-Nearest Neighbors model.
Secondly, we perform more detailed and extensive experiments to evaluate recommendation performance:
(i) we analyze the impact of the facets of trust on recommendation by tuning the components of MTM in a finer-grained way;
(ii) we integrate all the analyzed recommender systems in the same framework to uniformly evaluate the performance of algorithms;
(iii) we evaluate recommendation performance on new data (which was not previously used for training/validating the models) to comply with the application domains where models cannot be frequently optimized.

In the remainder of this paper, Section \ref{sec:questions} presents our research questions and outlines the experiments to answer them. Section \ref{sec:related} provides background concepts and positions our work in the related one. Sections \ref{model} and \ref{sec:MTM} present MTM and LOCABAL+. Section \ref{sec:datasets} describes the datasets used for the experiments and the instantiation of MTM on the available types of information. Section \ref{sec:validation} presents the validation methodology we used and the evaluation results. Section \ref{sec:discussion} discusses the evaluation results and outlines our future work. Section \ref{conclusions} concludes the paper.

\section{Research questions and experimental plan}
\label{sec:questions}
In the Multi-faceted Trust Model we integrate diverse facets of trust and, in a specific application domain, one or more of them might not be available or usable. Therefore, besides assessing their overall value in improving Top-N recommendation, we separately study their impact on recommendation performance.
We thus formulate the following research questions:
\begin{enumerate} 
\item[RQ1:] {\em Can multi-faceted trust be used to improve the performance of a trust-based recommender system with respect to only relying on social links and rating similarity among users?}
\item[RQ2:] {\em What is the impact of the multi-dimensional reputation of users, of the quality of their contributions, and of social links, on collaborative recommendation performance?}
\end{enumerate}
In order to answer these questions, we carry out experiments to measure the performance of LOCABAL+ on a spectrum of MTM configurations that tune in different ways the influence of the facets of trust we consider. We compare the performance of the algorithm when using or ignoring social links and trust statements about users and their contributions. 
We also compare the algorithm with the following baselines: the LOCABAL \citep{Tang-etal:13} and SocialMF \citep{Jamali-Ester:10} trust-based recommender systems based on Matrix Factorization; SVD++ \citep{Koren:08} and User-to-User Collaborative Filtering \citep{Desrosiers-Karypis:11} which only use rating similarity; a user-to-user Collaborative Filtering algorithm (henceforth denoted as U2USocial) that employs friend relations to estimate ratings in a K-Nearest Neighbors approach, instead of using rating similarity. 

We carry out the experiments on two large subsets the Yelp dataset \citep{Yelp-dataset}. The first one, Yelp-Hotel, concerns accommodation facilities; the second one, Yelp-Food, is focused on restaurants.

We evaluate Top-k recommendation performance of algorithms with k=10 by taking the rating scores of the dataset as ground truth. 
Following the recent trends in the evaluation of recommender systems described in \citep{Jannach-etal:16}, we measure their accuracy, error minimization, ranking capability and user coverage @k; see Section \ref{sec:metrics} for details.

\section{Background and related work}
\label{sec:related}

\subsection{Basic concepts: trust and reputation}
{\em Trust} is generally described as a positive expectation that an agent has about other agents' behavior, from a subjective perspective.
\cite{Gambetta:88} defines it as ``a particular level of the subjective probability with which an agent or group of agents will perform a particular action, both before [we] can monitor such action (or independently of [our] capacity of ever to be able to monitor it) and in a context in which it affects [our] own action''. 
Moreover, both \cite{Gambetta:88} and \cite{Goldbeck-Hendler:04} specify that a user trusts another one in a social network if (s)he believes that any future transaction with her/him will be rewarding rather than detrimental. On a more general perspective, \cite{Mui-etal:02} (and similarly \cite{Misztal:96}) elect ``subjective probability'' to subjective expectation, or degree of belief, to highlight that, more than a statistical probability, trust represents a belief status that an agent $A$ has about another agent $B$'s future behavior, given $B$'s past behavior and her/his {\em reciprocity} of action within a society.  

Different from trust, {\em reputation} describes a general ``expectation about an agent's behaviour based on information about or observations of its past behaviour'' \citep{Abdul-Rahman-Hailes:00}. Reputation has a global perspective and \cite{Mui-etal:02} describe it as the ``perception that an agent creates through past actions about its intentions and norms''. According to \cite{Misztal:96}, reputation ``helps us to manage the complexity of social life by singling out trustworthy people - in whose interest it is to meet promises''.

While the previously described works analyze trust and reputation from the global viewpoint of agent-to-agent interaction, a few ones contextualize it in online collaboration systems and social networks, which are the scope of our present work. 
Noticeably, \cite{McNally-etal:14} generalize trust relations by analyzing the occurrence of collaboration events that involve users; this makes it possible to link users because they have downloaded or bookmarked contents provided by other users, and so forth. These authors explain that reputation can derive from direct user-to-user interaction (e.g., when users are rated) or from indirect one; e.g., when they interact by virtue of some item. Moreover, it can derive from explicit trust statements, such as ratings, or from implicit ones like follower relations. 

Before describing the state of the art on trust-based recommender systems, it is worth briefly discussing the main issues affecting them.
Specifically, it may be questioned whether relying on social relations and global feedback about users is a safe approach to evaluate trustworthiness.
Some trust-based recommender systems focus on user-to-user relations and ignore the feedback on user actions because there is a general opinion that the latter could be biased. While we obviously agree that this may be true, we point out that {\em any} type of action that brings evidence about trust, including the establishment of friend relations, ratings, etc., could be performed with the aim of manipulating the reputation of some user. Therefore, data reliability assessment is a general pre-requisite for the development of recommender systems.
Indeed, the weaknesses of some models adopted in e-commerce and collaboration sites have been analyzed to suggest how to improve the robustness of Reputation Management Systems; e.g., see \citep{Resnick-Zeckhauser:02}. 
However, J{\o}sang et al. point out that these systems are challenged by strategic manipulation and by various types of attacks which cannot always be detected by statistical analyses \citep{Josang-etal:07,Josang-etal:09}. Therefore, \cite{Josang:12} ultimately highlights the importance of strengthening legislation as a barrier to discourage malicious behavior. 

\begin{table*}[t]
\centering 
\caption{Overview of the main social and trust-based recommender systems cited in the paper; ACO stands for Ant Colony Optimization.}
\resizebox{\columnwidth}{!}{%
{\def\arraystretch{1.5}
\begin{tabular}{lcccccc}
\hline
Citation & Algorithm & Technology & User filtering &
\multicolumn{1}{c}{
\renewcommand{\arraystretch}{0.7}
\begin{tabular}[c]{@{}c@{}}Social\\ 
regularization\end{tabular}}
 & Reputation  \\ 
\hline
\citep{Goldbeck-Hendler:06}  & FilmTrust & AVG & by trust & - & -  \\
\citep{Kuter-Goldbeck:07} & SUNNY & AVG & by trust & probabilistic & -  \\
\citep{DeMeo-etal:18}     & LGTR  & AVG & social proximity & - & by feedback \\
\citep{Parvin-etal:19}   & TCFACO  & AVG & by trust & ACO & -  \\
\citep{O'Donovan-Smyth:05}  & - & KNN & by trust & - & -  \\
\citep{Liu-Lee:10}  & - & KNN & - & - & -  \\
\citep{Moradi-Ahmadian:15}  & RTCF & KNN & \multicolumn{1}{c}{
\renewcommand{\arraystretch}{0.7}
\begin{tabular}[c]{@{}c@{}}by trust and\\ rating similarity\end{tabular}} & - & -  \\
\citep{Jamali-Ester:09}  & TrustWalker & Random Walk & by trust & - & -  \\
\citep{Deng-etal:14}  & RelevantTrustWalker & Random Walk & by trust relevancy & - & -  \\
\citep{Jamali-Ester:10}  & SocialMF & MF & - & rating-based & -  \\
\citep{Yang-etal:12}     & CircleCon & MF & by circle & - & -  \\
\citep{Tang-etal:13}     & LOCABAL  & MF & - & rating based & PageRank  \\
\citep{Guo-etal:15}     & TrustSVD  & MF & - & - & -  \\
\citep{Yang-etal:17}     & TrustMF  & MF & - & rating based & -  \\
\citep{Qian-etal:16}     & SoRS  & MF & - & rating based & 
\multicolumn{1}{c}{
\renewcommand{\arraystretch}{0.7}
\begin{tabular}[c]{@{}c@{}}by rating\\ conformity\end{tabular}}\\
\citep{Yuan-etal:11}     & MF.FM  & Collective MF & by group & rating based & -  \\
\citep{Ma-etal:11}     & SOREG  & MF & - & rating based & -  \\
\citep{Ma-etal:11b}     & RSTE  & PMF & - & rating based & -  \\
\citep{Liu-Aberer:13} & SoCo  & PMF & by context & - & -  \\
\citep{Du-etal:17} & SIACC  & Co-Clustering 
 & - & rating based & PageRank \\
 \hline
\end{tabular}
}}
\label{tab:algorithms}
\end{table*}

\subsection{Trust-based recommender systems} 
\label{sec:trust-based-RS}
The homophily \citep{McPherson-etal:01} and social influence \citep{Marsden-Friedkin:93} theories associate social links to user similarity. On this basis, social and trust-based recommender systems \citep{Richthammer-etal:17} exploit social networks as additional sources of information to complement rating data. These systems estimate user preferences by relying on the known social links existing between people; e.g., friend, follower and/or trust relations according to different inference techniques; see Table \ref{tab:algorithms}: 
\begin{itemize} 
\item
AVG: average rating of selected (e.g., trusted) social links \citep{Goldbeck-Hendler:04,Goldbeck-Hendler:06,Liu-Lee:10,Parvin-etal:19}.
\item 
KNN: K-Nearest Neighbors on social links \citep{O'Donovan-Smyth:05,Massa-Avesani:07,Groh-Ehmig:07,Moradi-Ahmadian:15,Ardissono-etal:17c}.
\item 
MF: Matrix Factorization (in some cases with Random Walk) on the matrices of ratings and social links \citep{Jamali-Ester:09,Jamali-Ester:10,Ma-etal:11,Yang-etal:12,Tang-etal:13,Deng-etal:14,Guo-etal:15,Yang-etal:17}.
\item 
PMF: Probabilistic Matrix Factorization on the matrices of ratings and social links \citep{Ma-etal:11b,Ma-etal:11c,Jiang-etal:12,Liu-Aberer:13,Chaney-etal:15}.
\item 
Probabilistic approaches on trust networks \citep{Kuter-Goldbeck:07,Li-etal:14}.
\item 
Co-clustering of ratings and trust matrices \citep{Du-etal:17}.
\end{itemize}
Differently, we generate personalized recommendations by relying on a compositional, multi-faceted trust model that includes complementary data about user behavior: i.e., not only local trust among users inferred from social relations, but also quality of user contributions (derived from the anonymous global feedback they receive) and multi-dimensional global reputation derived from diverse types of information, among which anonymous endorsements to user profiles. The integration of these facets of trust supports a rich computation of reputation based on complementary aspects of user behavior. Moreover, it makes it possible to compensate trust evidence in application domains in which some types of information are not available or cannot be used.
In particular, LOCABAL+ works with or without using social links. 

Some research about recommender systems studies the differences between trust and friends networks. \cite{Guo-etal:15}, \cite{Ma-etal:11} and \cite{Li-Fang:18} find out that, different from explicit trust relations (such as those among Epinions users \citep{Epinions-dataset}), friendship does not strictly imply preference similarity: user preferences are strongly correlated among trusted neighbors but they are only slightly positively correlated among ``trust-alike'' neighbors such as friends in social networks \citep{Guo-etal:15}.
Several authors recognize the importance of limiting the social context to the user's local proximity; for instance, \cite{Massa-Avesani:07} and \cite{Yuan-etal:11} prove that recommendation accuracy decreases when indirect social connections (i.e., paths of social links) are used to estimate user preferences. Moreover, \cite{Yang-etal:12} point out that users may trust different subsets of friends regarding different domains. In order to deal with this issue, authors propose various methods to filter the neighboorhood used for rating prediction; e.g., \cite{Yang-etal:17} use category-specific circles and \cite{Yuan-etal:11} use thematic groups to steer Matrix Factorization.
Moreover, KNN and AVG systems select neighbors by ranking the users directly linked to the current user in the social/trust network on the basis of their rating similarity \citep{Massa-Avesani:07,Liu-Lee:10,Li-etal:14,Moradi-Ahmadian:15,Ardissono-etal:17c,Parvin-etal:19}. Analogously, social regularization is used to increase the impact of like-minded users in Matrix Factorization: e.g., TrustMF \citep{Yang-etal:17} applies social regularization to users' direct social links and \cite{Yuan-etal:11} applies it to the members of thematic groups; RSTE \citep{Ma-etal:11b} and SOREG \citep{Ma-etal:11} integrate trust and rating similarity in Probabilistic Matrix Factorization, and \cite{Ma-etal:11c} use tag-based similarity to build a larger social context for regularization. Finally, SocialMF \citep{Jamali-Ester:10} employs rating similarity to regularize the impact of users who are reachable through a short path of social links in Random Walk.  
Other systems achieve similar filtering results by combining trust-based and item-based recommendation, as in TrustWalker \citep{Jamali-Ester:09}, or by filtering the users of the trust networks according to rating similarity, as in TCFACO \citep{Parvin-etal:19} and RelevantTrustWalker \citep{Deng-etal:14}. Finally, \cite{Du-etal:17} apply co-clustering to the matrices of ratings and social relations in order to identify like-minded users within social connections.

In comparison, we extend social regularization by tuning the impact of users on the Matrix Factorization process on the basis of both rating similarity and global multi-dimensional reputation. In other words, we select neighbors for rating prediction by privileging users who are trustworthy {\em and} like-minded. This approach improves prediction accuracy because it enhances the quality of the rating information used to estimate preferences.

Building on social influence theories, \cite{Guo-etal:15} propose TrustSVD that extends SVD++ \citep{Koren:08} to jointly factorize the rating and trust matrices: they learn a truster model that describes how people are influenced in item evaluation by their parties's opinions. TrustMF \citep{Yang-etal:17} learns both the truster and trustee models to consider the fact that, in a social network, people mutually influence each other. \cite{Jiang-etal:12} investigate the relation between social influence and personal preferences.
Finally, some researchers leverage the local and global perspectives of social influence building on the observation that in the physical world humans ask for opinions from both local friends and highly reputable people; e.g., \cite{Qian-etal:16,Tang-etal:13,Hu-etal:18}. Specifically, in LOCABAL, \cite{Tang-etal:13} combine rating similarity and social links with users' global reputation, which is based on the PageRank \citep{Page-etal:99} score as a measure of importance in the social network. 

We use both local trust and global reputation to steer personalized recommendation. However, we propose a multi-faceted trust model to determine users' trustworthiness on the basis of complementary types of information about their behavior, including social relations and global feedback from other users. As previously noticed, this makes it possible to tune Matrix Factorization on the basis of the contributions of trustworthy, like-minded people, using public, anonymous information.
To the best of our knowledge, the only other work that employs global feedback about users is LGTR by \cite{DeMeo-etal:18}, which defines global reputation on the basis of the feedback collected by user actions and of a local context depending on social relations. However, in neighbor identification, LGTR discards rating similarity, which is very useful to select like-minded users for preference prediction. Moreover, LGTR does not take review quality and endorsements to user profiles into accout. Our model is thus more general than this one.

Some trust-based recommender systems assume the existence of both positive and negative evidence about trust as, e.g., in the social networks where users can rate other users positively or negatively  \citep{Li-etal:11,Victor-etal:11,Rafailidis-Crestani:17}. In our work, we start from the consideration that the trust models provided by several social networks are only based on the expression of ``likes''. Therefore, we propose a model that can also work on positive-only feedback to comply with them.

Some works associate users' reputation to rating conformity; e.g., 
\linebreak 
\cite{O'Donovan-Smyth:05} and \cite{Li-etal:13} base reputation on the percentage of ratings provided by a user that agree with those of the other people. \cite{Su-etal:17} cluster users on the basis of rating similarity and consider the largest cluster as the ``honest'' group. In the SoRS recommender system \cite{Qian-etal:16} derive reputation by iteratively calculating the correlation of the historical ratings provided by a user and the quality of items emerging from the rating scores they receive.
However, review conformity does not fully characterize quality; e.g., \cite{Victor-etal:11} point out that controversial reviews must be considered and matched to individual preferences. 
We thus leave this aspect for our future work.

\section{Multi-faceted Trust Model (MTM)}
\label{model}
MTM is aimed at computing users' trustworthiness in the context of recommender systems. It integrates local trust between users (inferred from social relations, in line with trust-based recommender systems research) with the public, anonymous feedback received by users and by their contributions in a social network. MTM is compositional and supports the inclusion or exclusion of facets of trust to comply with the requirements of the application domain of interest.
We identified the classes of evidence about trust of MTM by analyzing the information publicly provided by social networks and e-commerce sites such as \cite{Yelp}, \cite{Booking}, \cite{Expedia}, \cite{LibraryThing}, \cite{Amazon}, Ciao \citep{CIAO-DVD} and \cite{Epinions-dataset}. However, we generalized those indicators to enhance the applicability of our model to heterogeneous domains. 
In the following we describe each class in detail.

\subsection{Quality of individual contributions on an item}
\label{sec:classC}

By individual contribution on an item we mean a piece of information that a user provides about it. A contribution is usually a review associated with a rating score but we describe our model at a more general level because in some online services users can post different types of content; e.g., the Yelp social network allows to write both reviews and tips about items.

Let ${\cal U}$ be the set of users and ${\cal I}$ the set of items of a service. Given $v \in {\cal U}$ and $i \in {\cal I}$ we denote an individual contribution provided by $v$ on $i$ as $contr_{vi}$.
Then, we define the quality of $contr_{vi}$ ($fContr_{vi}$, in [0, 1]) on the basis of the amount of feedback that $contr_{vi}$ has received. We measure quality in a relative way with respect to the most popular contributions on the same item in order to be robust with respect to item-specific biases:
\begin{itemize}
    \item 
    In the social networks that provide both positive and negative feedback about contributions we take inspiration from the definition of gold standard helpfulness of reviews defined in \citep{Kim-etal:06,Raghavan-etal:12,O'Mahony-Smyth:18}. In those works, helpfulness is defined as the ratio between the number of positive evaluations ($positiveVotes_{contr_{vi}}$) and the total number of evaluations 
    \linebreak ($positiveVotes_{contr_{vi}}+negativeVotes_{contr_{vi}}$) that a contribution receives:
    \begin{equation}
        helpfulness_{contr_{vi}} = \frac{positiveVotes_{contr_{vi}}}{positiveVotes_{contr_{vi}}+negativeVotes_{contr_{vi}}}
        \label{eq:Kim}
    \end{equation} 
    We define the quality of $contr_{vi}$ in a relative way with respect to the best contribution on the same item as follows:
     \begin{equation}
        fContr_{vi} = \frac{helpfulness_{contr_{vi}}}{\max\limits_{a\in {\cal U}}helpfulness_{contr_{ai}}}
        \label{eq:positiveNegativeFeedback}
    \end{equation} 
    \item
    In the social networks that only support positive feedback we compute quality as the ratio between the overall number of appreciations obtained by $contr_{vi}$ ($appreciations_{contr_{vi}}$) and the maximum number of appreciations received by the other contributions on the same item:
    \begin{equation}
       fContr_{vi} = 
            \frac{appreciations_{contr_{vi}}}
            {\max\limits_{a\in {\cal U}}appreciations_{contr_{ai}}} 
        \label{eq:fPositiveFeedback}
    \end{equation}
\end{itemize}

\subsection{Multi-dimensional global reputation}
\label{sec:MGR}
We define {\em multi-dimensional global reputation} building on heterogeneous types of information about users to capture different aspects of their behavior.
Let ${\cal U}$ be the set of users, ${\cal I}$ the set of items, $v\in {\cal U}$ and $i \in  {\cal I}$. We define the following sub-classes of trust evidence: 
\begin{enumerate}
    \item[(P)]
    {\em Importance of the user in the social network} ($imp_v$ in [0, 1]), based on her/his social connections. Similar to LOCABAL, we use PageRank \citep{Page-etal:99} to model this type of indicator. PageRank estimates the relative importance of nodes in a graph by counting the number and quality of the links that enter them, under the assumption that being referenced by others is a quality sign. We compute $imp_v$ as:
    \begin{equation}
        imp_v = \frac{1} 
            {1+log(rank_v)}
        \label{eq:PageRank}
    \end{equation} 
    where $rank_v \in [1, |U|]$ is the PageRank score of $v$ and the most important user is ranked with 1; see \citep{Tang-etal:13}.
    \item[(U)] 
    {\em Global feedback about the user's profile:}
        \begin{itemize}
         \item {\em User profile endorsements and public recognition} ($fEndors_v$, in [0, 1]). This class represents the global types of feedback that user profiles receive from the social network. It may have different instances representing individual trust indicators. We consider the appreciations that a user {\em v} receives from the other members of ${\cal U}$ (e.g., ``likes''), public assessments of reputation which some social networks grant to their best contributors, and the number of friends, fans, or followers in the social networks that disclose the number but not the identity of users. Similar to the evaluation of the feedback on user contributions, we compute the value of each trust indicator as the ratio between the number of appreciations received by {\em v}, denoted as $appreciations_v$, and the maximum number of appreciations received by a user $a \in {\cal U}$:
        \begin{equation}
            fEndors_v = \frac{appreciations_v} 
            {\max\limits_{a \in {\cal U}} appreciations_a}
        \label{eq:fEndors}
        \end{equation} 
        In this way we are able to assign a value that indicates the importance of each user profile with respect to the profiles of the other users of the community, on the basis of public or anonymous data.
        \item {\em Visibility} ($vis_v$ in [0, 1]). This class is aimed at estimating how popular $v$ becomes, thanks to her/his contributions. Intuitively, the visibility describes the impact of the user's contributions in the social network as observed from the feedback they receive. We compute $vis_v$ as the ratio between the number of appreciations received by $v$ and the total number of contributions provided by her/him, normalized by the maximum number of appreciations acquired by the other members of ${\cal U}$:
        \begin{equation}
            vis_v = \frac{appreciations_v}
            {\max\limits_{a \in {\cal U}} appreciations_a*|Contributions_v|}
        \label{eq:visibility}
        \end{equation} 
        where $Contributions_v$ is the set of contributions authored by $v$.
        \end{itemize}
    \item[(Q)] 
    {\em Quality of the user as a contributor} ($q_v$, in [0, 1]) with respect to the other members of ${\cal U}$. 
    
    This class is aimed at providing an overall evaluation of the user by considering the feedback received by all her/his contributions. As in the previous cases, we compute quality in a relative way with respect to the best contributor of the social network. Specifically:
    \begin{itemize}
    \item
    In the social networks that only provide positive feedback about contributions we define $q_v$ as follows: 
    \begin{equation}
            q_v = \frac {\sum\limits_{c_1 \in Contributions_v} appreciations_{c_1}}
            {\max\limits_{a \in {\cal U}} \sum\limits_{c_2 \in Contributions_a} appreciations_{c_2}}
        \label{eq:overallQuality}
    \end{equation} 
    where $appreciations_{c_1}$ is the number of appreciations received by contribution $c_1$ and $c_1$ is authored by user $v$ (analogously for $c_2$).
    \item
    In the social networks that provide positive and negative feedback about reviews, we apply Equation \ref{eq:overallQuality} by replacing $appreciations_{c_n}$ with $fContr_{c_n}$ computed according to Equation \ref{eq:positiveNegativeFeedback}; i.e., we compute the relative quality of the user's contributions by taking both the positive and negative votes they receive into account.
    \end{itemize}
\end{enumerate}
The previously described classes of trust evidence are generic and most of them could be mapped to multiple indicators. For instance, Yelp supports different types of endorsements to user profiles, such as ``thanks'' and ``Elite'' recognition. In other cases, the social relation between users might be mapped to friends, follower and trust links. 
In order to obtain a single value representing a user $v$'s multi-dimensional reputation, we fuse these indicators by computing their average, assuming that they additively contribute to increasing {\em v}'s trustworthiness. 
Let's consider a set ${\cal F}$ of indicators that are instances of the P, U and Q classes. We define the {\em multi-dimensional global reputation} of {\em v}, denoted as $mgr_v$ (in [0, 1]) as:
\begin{equation}
     mgr_{v} = 
     \frac{
     \sum _{{l=1}}^{|{\cal F}|} C_l * indicator_l}
     {\sum _{{l=1}}^{|{\cal F}|} C_l }
    \label{eq:multi-dimensional-rep}
\end{equation}
where $C_l$ can be set to 1 to take the trust indicator (e.g., social links) into account, 0 otherwise.
We assume that each indicator is computed according to the method defined for the class to which it belongs.

It is worth noting that in Equation \ref{eq:multi-dimensional-rep} all the indicators have the same weight because, for simplicity, we assume that they equally contribute to {\em v}'s reputation. In our future work, we plan to carry out a deeper analysis to understand the impact of different weighting schemes on recommendation performance. For this purpose, we will carry out experiments with LOCABAL+ by setting these weights to different values in [0, 1].

\subsubsection{Multi-faceted trust}
The multi-faceted trust, $mft_{vi}$ (in [0, 1]), describes the overall trust in the rating provided by a user $v$ on an item $i$, given {\em v}'s multi-dimensional reputation and the quality of her/his contribution about $i$. We use multi-faceted trust values in LOCABAL+ to tune the influence of rating scores in the Matrix Factorization process used to learn the latent user and item vectors; see Section \ref{sec:LOCABAL+}. We define $mft_{vi}$ as follows:
    \begin{equation}
        mft_{vi} = \beta * mgr_v + C (1-\beta) fContr_{vi}
        \label{eq:mt}
    \end{equation}
where
\begin{itemize}
    \item 
    $mgr_{v}$ is $v$'s multi-dimensional reputation; see Equation \ref{eq:multi-dimensional-rep}.
    \item 
    $fContr_{vi}$ is the quality of the contribution provided by $v$ on item $i$ and is computed according to Equations \ref{eq:positiveNegativeFeedback} or \ref{eq:fPositiveFeedback}, depending on the type of feedback (positive/negative) that contributions can receive.
    \item 
    $\beta$ takes values in the [0, 1] interval and balances the relative weight of $mgr_v$ and $fContr_{vi}$ in the computation of $mft_{vi}$. The higher $\beta$, the stronger is the impact of multi-dimensional reputation on trust. 
    \item 
    $C$ can be either 0 or 1 and is used to ignore or use the feedback on contributions in the evaluation of $mft_{vi}$; by default, $C=1$.
 \end{itemize}
As discussed below in Section \ref{sec:validation}, the best configuration of the $\beta$ parameter in Equation \ref{eq:mt} depends on the dataset to which the trust model has to be applied and it can be empirically found by using the MTM model within a recommender system (LOCABAL+ in our case) and checking its performance. In the datasets we have considered, the best values are somehow low (e.g., 0.1 or 0.3), which means that the global feedback on user contributions, represented by $fContr_{vi}$, is very useful to steer recommendation.

\section{Recommendation model}
\label{sec:MTM} 

We describe LOCABAL+ incrementally, starting from the main concepts that characterize the LOCABAL trust-based recommender system.

\subsection{Basic Collaborative Filtering with Matrix Factorization}
\label{sec:MF}
Basic Collaborative Filtering builds on the assumption that, if people rated items similarly in the past, they will do it again in the future. Thus, it uses rating similarity in preference estimation. The algorithms based on Matrix Factorization assume that a few latent patterns influence rating behavior and they perform a low-rank matrix factorization on the users-items rating matrix; e.g., see SVD++ \citep{Koren:08}. Given the following notation:
\begin{itemize} 
\item  
${\cal U} = \{u_1, \dots, u_n\}$ is the set of users and
${\cal I} = \{i_1, \dots, i_m\}$ is the set of items.
\item 
{\bf R} $\in {\rm I\!R}^{n \times m}$ is the users-items rating matrix.
\item 
${\bf R}_{xy}$ is the rating score given by user $u_x \in {\cal U}$ to item $i_y \in {\cal I}$, if any:
\begin{itemize}
    \item 
    ${\cal O} = \{<u_x, i_y> |~ {\bf R}_{xy} \neq 0 \}$ is the set of known ratings (ground truth)
    \item
${\cal T} = \{<u_x, i_y> |~ {\bf R}_{xy} = 0 \}$ is the set of unknown ratings
\end{itemize}
\end{itemize}
Assuming $K$ latent factors, {\bf u}$_x \in {\rm I\!R}^K$ denotes the user preference vector of $u_x$ and {\bf i}$_y \in {\rm I\!R}^K$ denotes the item characteristic vector of $i_y$.

In order to learn these vectors, the recommender system solves the following optimization problem:
\begin{equation}
    \min\limits_{{\bf U}, {\bf I}} \sum\limits_{<u_x, i_y> \in {\cal O}} 
     ({\bf R}_{xy} - {\bf u}_x^T {\bf i}_y)^2 +
     \lambda(||{\bf U}||^2_ F + ||{\bf I}||^2_F)
    \label{eq:MatrixFactorization}
\end{equation}
where 
\begin{itemize} 
\item 
${\bf U} = [{\bf u}_1, \dots, {\bf u}_n] \in {\rm I\!R}^{K \times n}$ and
${\bf I} = [{\bf i}_1, \dots, {\bf i}_m] \in {\rm I\!R}^{K \times m}$.
\item 
$||.||_F$ denotes the Frobenius Norm and $||{\bf U}||^2_ F + ||{\bf I}||^2_F$ are the regularization terms to avoid over-fitting.
\item 
$\lambda>0$ controls the impact of {\bf U} and {\bf I} on regularization.
\end{itemize}

\subsection{LOCABAL}
\label{sec:LOCABAL}
LOCABAL \citep{Tang-etal:13} extends Collaborative Filtering based on Matrix Factorization in two ways:
\begin{enumerate} 
\item 
It exploits a user's local social context to learn her/his preference vector by considering both rating similarity and social relations, regularized on the basis of the former. In this way, rating estimation can benefit from the contribution of users who are socially related to the current user but, at the same time, have similar preferences as her/him.
\item
It relies on the user's global social context, represented by her/his reputation, to weight the contribution of rating similarity in Matrix Factorization. 
Global reputation is computed using PageRank as described in Section \ref{sec:MGR}, Equation \ref{eq:PageRank}, page \pageref{eq:PageRank}.
\end{enumerate}
In detail:
\begin{itemize} 
    \item 
    Let {\bf T} $\in {\rm I\!R}^{n \times n}$ be the users-users social relation matrix. ${\bf T}_{uz}\neq 0$ denotes the existence of a direct social link between $u_x \in {\cal U}$ and $u_z \in {\cal U}$. Zero values mean that users are not socially related.
    \item 
    Let ${\cal N}_x = \{u_z ~|~ {\bf T}_{xz} = 1 \}$ be the set of $u_x$'s direct social links.
\item
Let {\bf S}$ \in {\rm I\!R}^{n \times n}$ be a users-users trust matrix whose cells represent the strength of the social relations between users, depending on their rating similarity. For $u_z\in {\cal N}_x$, {\bf S}$_{xz} = \sigma(u_x, u_z)$, where $\sigma(u_x, u_z)$ is the Cosine similarity of $u_x$ and $u_z$'s rating vectors.
\end{itemize}
LOCABAL solves the following optimization problem:
\begin{equation}
    \begin{split}
    \min\limits_{{\bf U}, {\bf I}, {\bf H}} \sum\limits_{<u_x, i_y> \in {\cal O}} 
     w_x ({\bf R}_{xy} - {\bf u}_x^T {\bf i}_y)^2 +
     \alpha \sum\limits_{x=1}^{n} \sum\limits_{u_z \in N_x} ({\bf S}_{xz} - {\bf u}_x^T {\bf H u}_z)^2 + \\
     \lambda(||{\bf U}||^2_ F + ||{\bf I}||^2_F  + ||{\bf H}||^2_F)
    \end{split}
    \label{eq:LOCABAL}
    \end{equation}
where
\begin{itemize}
    \item 
    $w_x$ in [0, 1] is $u_x$'s global reputation computed by applying Equation \ref{eq:PageRank}. This weight tunes the contribution given by rating similarity so that highly reputable users influence the Matrix Factorization process more strongly than the other ones.
    \item 
    $\alpha >=0$ tunes the contribution given by $u_x$'s local social context.
    \item 
    {\bf H}$\in {\rm I\!R}^{K \times K}$ captures user preference correlation: if $u_x$ and $u_z$ are strongly connected in {\bf S}$_{xz}$, then their preferences should be tightly correlated via {\bf H}.
    We remind that $K$ is the number of latent factors.
    \item 
    $\lambda>=0$ controls the impact of {\bf U}, {\bf I} and {\bf H} on regularization.
\end{itemize}
As discussed in Section \ref{sec:related}, this algorithm estimates trust by relying on social links. If this information is not available, LOCABAL cannot be applied or it reduces to SVD \citep{Koren-etal:09}, by setting $w_x = 1$ and ignoring the trust matrix that cannot be computed. Our MTM model is aimed at providing a more general solution, which can be applied to complementary types of evidence about trust as can be found in social networks.

\subsection{LOCABAL+}
 \label{sec:LOCABAL+}

LOCABAL+ extends LOCABAL in two ways: 
\begin{enumerate} 
\item 
It models global context by taking multi-faceted trust into account.
\item 
It tunes social regularization on the basis of both rating similarity and multi-dimensional global reputation. 
\end{enumerate}
We consider the following optimization problem to be solved in order to learn the user preference and item characteristic vectors: 
    \begin{equation}
    \begin{split}
    \min\limits_{{\bf U}, {\bf I}, {\bf H}} \sum\limits_{<u_x, i_y> \in {\cal O}} 
     mft_{xy} ({\bf R}_{xy} - {\bf u}_x^T {\bf i}_y)^2 +
     \alpha \sum\limits_{x=1}^{n} \sum\limits_{u_z \in N_x}
     mgr_z ({\bf S}_{xz} - {\bf u}_x^T {\bf H u}_z)^2 \\
     + \lambda(||{\bf U}||^2_ F + ||{\bf I}||^2_F  + ||{\bf H}||^2_F)
     \end{split}
    \label{eq:LOCABAL+}
    \end{equation}
    where:
    \begin{itemize} 
    \item 
    $mft_{xy}$ represents the multi-faceted trust towards user $u_x$ in the context of item $i_y$; see Equation \ref{eq:mt} in page \pageref{eq:mt}. 
    This weight tunes the estimation of ratings in the Matrix Factorization process by taking users' global reputation and quality of contributions into account; i.e., by looking at users from a broad perspective on their behavior.
    We assume that we can estimate missing ratings more precisely by giving more importance to the ratings authored by users whose multi-faceted trust is high.
    \item 
    {\bf S} $\in {\rm I\!R}^{n \times n}$ is a users-users trust matrix such that, for $u_z\in {\cal N}_x$, 
    \linebreak {\bf S}$_{xz}$ is set to the Pearson Correlation similarity ($PC$) of $u_x$ and $u_z$'s rating vectors, limited to the set of items rated by both users:
    \begin{equation}
        PC(u_x, u_z) = 
        \frac{\sum\limits_{i_y \in {\cal I}_{xz}} (r_{xy}-\bar{r}_x)(r_{zy}-\bar{r}_z)}
        {\sqrt{
            \sum\limits_{i_y \in {\cal I}_{xz}}(r_{xy} - \bar{r}_{x})^2
            \sum\limits_{i_y \in {\cal I}_{xz}}(r_{zy} - \bar{r}_{z})^2}
        }
        \label{eq:Pearson}
    \end{equation}
    where ${\cal I}_{xz}$ is the set of items rated by both $u_x$ and $u_z$, $r_{xy}$ is the rating given by $u_x$ to $i_y$ (and analogously for $r_{zy}$), $\bar{r}_x$ ($\bar{r}_z$) is the mean value of $u_x$'s ($u_z$'s) ratings.
    
    As suggested in \citep{Ricci-etal:11}, we use Pearson Correlation similarity, instead of the Cosine similarity used in LOCABAL, because the latter does not consider the differences in the mean and variance of the ratings made by $u_x$ and $u_z$. Pearson similarity removes the effects of mean and variance.
    \item 
    $mgr_z$ is the multi-dimensional global reputation of $u_z$ and tunes preference correlation in the {\bf H} matrix (which depends on rating similarity, given {\bf S}) on the basis of $u_z$'s multi-dimensional global reputation.
    By adding the $mgr_z$ factor we impose that, the more reputable are $u_x$'s friends, the higher impact they have in the estimation of their own similarity with $u_x$. Therefore, highly trustworthy users influence social regularization more than the others.
    \end{itemize}
As $mft_{xy}$ and $mgr_z$ are based on a compositional model, they can be computed by using a subset of the trust facets considered so far; e.g., by ignoring social links, feedback on user profiles or feedback about user contributions. In those cases, LOCABAL+ runs with a lower amount of information about users but it can still work as a trust-based recommender system. Specifically, in the experiments we carried out, the algorithm reaches satisfactory performance results also with partial evidence about trust; see Section \ref{sec:validation}.

Obviously, the flexibility of MTM comes with a cost, i.e., the effort needed to map the facets of trust to the types of information available in the application domain in which the recommender system is used. This effort consists of understanding the semantics of the evidence about trust (e.g., types of feedback that are provided by users) and choosing the corresponding classes of trust in MTM. However, as previously discussed, we defined these classes by analyzing several social networks to abstract from the particular types of information they offer and to model trust in a general way. The next section describes the datasets we used for our experiments and the mappings we defined to apply LOCABAL+ to these datasets. Moreover, it sketches the work that should be done to map MTM to a different type of social network in order to give the reader a broader idea of the work to be done.

\section{Datasets}
\label{sec:datasets}

For our experiments we use two subsets of the \citep{Yelp-dataset} dataset to analyze data about user behavior in different domains: accommodations versus restaurants. 

The Yelp dataset contains information about the users of the social network and about a large set of businesses including food, accommodation, transportation, health, education and so forth. Yelp members can establish bidirectional friend relations to share posts; moreover, they can establish stricter unidirectional fan relations to get access to the contributions provided by other users.
The dataset is structured as follows:
\begin{itemize}
    \item 
    Each item (business) is associated with a list of tags representing the categories to which it belongs; e.g., a restaurant might be associated with the ``Indian'' tag to specify the type of cuisine it offers.
    The full list of Yelp categories is available at
    \newline 
    {\footnotesize \url{https://www.yelp.com/developers/documentation/v3/category_list}}.
    \item
    Each item is associated with the rating scores and with textual reviews and tips provided by the members of Yelp. Every user can post a contribution (including review+rating, and possibly tip) on the same item. Item ratings take values in a [1,5] Likert scale where 1 is the worst value and 5 is the best one.
    \item 
    User contributions are associated with the appreciations they receive from Yelp members; i.e., ``useful'', ``funny'' and ``cool'' for reviews, ``like'' for tips.
    \item 
    The dataset publishes friend relations but it only provides the number of fans of each user. Therefore, only the former data can be used to infer direct {\em trust-alike} relations among users; see Section \ref{sec:trust-based-RS}.
    \item 
    The dataset publishes various types of endorsement that user profiles can receive: e.g., every year Yelp rewards its most valuable contributors by attributing them the status of Elite users. Moreover, each user profile can receive {\em compliments} by other Yelp users; e.g., ``write more'', ``thanks'' and ``great writer''. 
\end{itemize}
Notice that both compliments and appreciations represent positive feedback about users and contributions; moreover, the dataset reports the number of compliments and appreciations but not the identities of the people who provided them. This type of feedback thus represents an important anonymous source of trust information that can be used by a recommender system.

\begin{table}[t]
\centering
\caption{Statistics about the Yelp-Hotel dataset.}
{\footnotesize
\begin{tabular}{lcccc}
\hline\noalign{\smallskip}
Measure & & & &  Value \\ 
\hline\noalign{\smallskip}
\#Users & & & &  654 \\
\#Items (businesses) & & & & 1081\\
\#Ratings & & & & 10081\\
\#Friend relations & & & & 11554\\
Sparsity of users-items rating matrix & & & &  0.9857 \\
Sparsity of users-users friends matrix & & & &  0.9729 \\
\hline\noalign{\smallskip}
Measure  & Min & Max & Mean & Median \\ 
\hline\noalign{\smallskip}
\#Elite years of individual users           & 0 & 13    & 4.4052   & 4 \\
\#Compliments received by individual users  & 0 & 45018 & 724.1177 & 110 \\
\#Fans of individual users                  & 0 & 1803  & 91.9740  & 41 \\
\#Appreciations on reviews provided by ind. users  & 0  & 5194 & 112.7064  & 53  \\
\#Appreciations on tips provided by ind. users   & 0  & 152  & 2.5107  & 0  \\
\#Appreciations received by individual reviews         & 0  & 559  & 3.9897  & 3  \\
\#Friends of individual users                          & 0  & 224  & 17.6667 & 7 \\
\hline
\end{tabular}%
}
\label{t:Yelp-Hotel-statistics}
\end{table}

\subsection{Yelp-Hotel}
\label{sec:Yelp-Hotel}
Yelp-Hotel is obtained by filtering the complete Yelp dataset on users who provided at least 10 ratings and on businesses tagged with at least one category associated with accommodation facilities. The tags used to filter the dataset are:
Hotels,
Mountain Huts,
Residences,
Rest Stops,
Bed \& Breakfast,
Hostels,
Resorts.

Table \ref{t:Yelp-Hotel-statistics} provides information about this dataset. It can be noticed that user profiles receive various types of feedback; e.g., the median number of Elite years is 4 and the median number of compliments to user profiles is 110. Also anonymous fans contribute to global reputation (median = 41). Moreover, the dataset contains a relatively high amount of feedback about user contributions: the median number of appreciations is 53 for reviews and 0 for tips. The number of compliments, fans, appreciations, etc. reaches very high values in some cases: for each type of feedback, the distribution of individuals (users or contributions) has a long tail. Both the users-items and the users-users friends matrices are sparse.

\begin{table}[t]
\centering
\caption{Statistics about the Yelp-Food dataset.}
{\footnotesize
\begin{tabular}{lcccc}
\hline\noalign{\smallskip}
Measure & & & &  Value \\ 
\hline\noalign{\smallskip}
\#Users & & & &  8432 \\
\#Items (businesses) & & & & 8157\\
\#Ratings & & & & 198759\\
\#Friend relations & & & & 160891\\
Sparsity of users-items rating matrix & & & &  0.9971
\\
Sparsity of users-users friends matrix & & & &  0.9977
\\
\hline\noalign{\smallskip}
Measure  & Min & Max & Mean & Median \\ 
\hline\noalign{\smallskip}
\#Elite years of individual users  & 0            & 13            & 1.4154
        & 0 \\
\#Compliments received by individual users     & 0            & 24635
        & 55.0733

       & 4 \\
\#Fans of individual users                                             & 0            & 1803         & 10.1950
      & 2 \\
\#Appreciations on reviews provided by ind. users  & 0            & 9023
        & 81.4163

      & 27  \\
\#Appreciations on tips provided by ind. users                    & 0            & 154          & 0.3876

       & 0  \\
\#Appreciations received by individual reviews & 0            & 642         & 3.4539
        & 1  \\
\#Friends of individual users & 0 & 1231 & 19.0810

 & 4 \\
 \hline
\end{tabular}%
}
\label{t:Yelp-Food-statistics}
\end{table}

\subsection{Yelp-Food}
\label{sec:Yelp-Food}
Yelp-Food, about 10 times larger than Yelp-Hotel, is obtained by filtering the complete Yelp dataset on users who provided at least 10 ratings and on businesses located in the cities of Phoenix, Toronto, Pittsburgh which are tagged with at least one category describing a type of restaurant (e.g., ``Indian'' and ``Italian'') for a total of 85 categories.\footnote{
The selection of businesses to define the Yelp-Food is based on the following tags:
American,
Argentine,
Asian Fusion,
Australian,
Austrian,
Bangladeshi,
Belgian,
Brasseries,
Brazilian,
British,
Cambodian,
Cantonese,
Catalan,
Chinese,
Conveyor Belt Sushi,
Cuban,
Czech,
Delis,
Empanadas,
Falafel,
Filipino,
Fish \& Chips,
French,
German,
Greek,
Hawaiian,
Himalayan/Nepalese,
Hot Pot,
Hungarian,
Iberian,
Indian,
Indonesian,
Irish,
Italian,
Japanese,
Japanese Curry,
Korean,
Latin American,
Lebanese,
Malaysian,
Mediterranean,
Mexican,
Middle Eastern,
Modern European,
Mongolian,
New Mexican Cuisine,
Noodles,
Pakistani,
Pan Asian,
Persian/Iranian,
Peruvian,
Piadina,
Pizza,
Poke,
Polish,
Polynesian,
Portuguese,
Ramen,
Russian,
Salad,
Scandinavian,
Scottish,
Seafood,
Shanghainese,
Sicilian,
Singaporean,
Soup,
Southern,
Spanish,
Sri Lankan,
Steakhouses,
Sushi Bars,
Syrian,
Tacos,
Tapas Bars,
Tapas/Small Plates,
Teppanyaki,
Tex-Mex,
Thai,
Turkish,
Ukrainian,
Vegan,
Vegetarian,
Vietnamese,
Wraps.
} 

As shown in Table \ref{t:Yelp-Food-statistics}, the median number of compliments, fans and appreciations is very low and it reaches higher values only for the number of appreciations on reviews provided by individual users. Moreover, for each type of feedback, the distributions of users and reviews have long tails.

\subsection{Trust indicators for both datasets}

Let ${\cal U}$ and ${\cal I}$ be the sets of users and items of the dataset; let $u, v \in {\cal U}$ and $i \in {\cal I}$. We define the following trust indicators:
\begin{itemize}
\item {\em Quality of individual contributions on an item } ($fContr_{vi}$, in [0, 1]).
    
For this class of trust evidence we apply Equation \ref{eq:fPositiveFeedback}, which is suitable for positive-only feedback. We map $appreciations_{contr_{vi}}$ to the possibly different types of feedback that a contribution $contr_{vi}$ can receive:
    \begin{itemize} 
    \item
     $appreciations_{contr_{vi}} = useful_{contr_{vi}}+fun_{contr_{vi}}+cool_{contr_{vi}}$ for item reviews
     \item 
     $appreciations_{contr_{vi}} = like_{contr_{vi}}$ for tips
     \end{itemize} 
where $useful_{contr_{vi}}$ is the number of ``useful'' appreciations received by $contr_{vi}$ (a review), $fun$ is a shortener for ``funny'' and so forth.
    \item 
    {\em Multi-dimensional global reputation} ($mgr_v$ in [0, 1]).
    
    Following the approach described in Section \ref{sec:MGR}, we compute the multi-dimensional global reputation of a user $v$ by fusing in Equation \ref{eq:multi-dimensional-rep} the indicators described in the remainder of this section:
    \begin{equation}
    mgr_v = \frac{C_1imp_v+C_2elite_v+C_3lup_v+C_4opLeader_v+C_5vis_v+C_6q_v}
    {C_1+C_2+C_3+C_4+C_5+C_6}
    \label{eq:mgr_Yelp}
    \end{equation} 
        \begin{enumerate}
        \item[(P)] {\em Importance of the user in the social network} ($imp_v$ in [0, 1]).
    
        In order to compute the PageRank score of users, we transform each bidirectional friend relation into two unidirectional social links. In this way, we can apply the approach described in Section \ref{sec:MGR} and Equation \ref{eq:PageRank} to compute reputation on the basis of the connections among the users of the social network.
        \item[(U)] 
        {\em Global feedback on the user's profile}. 
    
        We consider the following trust indicators:
        \begin{itemize}
        \item $elite_v$ (in [0, 1]). 
        We map the number of years in which $v$ has the Elite status to $appreciations_v$ in Equation \ref{eq:fEndors}:
        \begin{equation}
            elite_v = \frac{\#EliteYears_v}{\max\limits_{a \in {\cal U}}\#EliteYears_a}
        \end{equation} 
        \item
        $lup_v$ (degree of liking of user profile, in [0, 1]). We map the number of compliments (``more'', ``thanks'' - $thks$, ``great writer'' - $gw$) received by $v$ to
        $appreciations_v$ in Equation \ref{eq:fEndors}:
        \begin{equation}
            lup_v = \frac{more_v + thks_v + gw_v} 
            {\max\limits_{a \in {\cal U}} (more_a + thks_a + gw_a)}
        \end{equation} 
        where $more_v$ is the number of ``more'' compliments received by $v$, and similar for the other variables.
        \item 
        $opLeader_v$ (opinion leader degree, in [0, 1]). The number of anonymous fans of a user $v$, $fans_v$, can be interpreted as a global recognition of her/his profile. We thus map this number to $appreciations_v$ in Equation \ref{eq:fEndors}: 
        \begin{equation}
            opLeader_v = \frac{fans_v}
            {\max\limits_{a \in {\cal U}} fans_a}
        \end{equation}
        \item
        $vis_v$ (visibility, in [0, 1]). We map the number of compliments received by $v$ to $appreciations_v$, and the reviews and tips ($Revs_v \cup Tips_v$) authored by $v$ to $Contributions_v$ in Eq. \ref{eq:visibility}:
        \begin{equation}
            vis_v = \frac{more_v + thks_v + gw_v}
                    {\max\limits_{a \in {\cal U}} (more_a + thks_a + gw_a)|Revs_v \cup Tips_v|}
        \end{equation}
        \end{itemize}
    \item[(Q)] 
    {\em Quality of the user as a contributor} ($q_v$ in [0, 1]).
    
    We assume that the quality of a contributor depends on both the reviews and tips authored by her/him. Therefore, for this indicator, we map $Contributions_v$ to the reviews and the tips provided by $v$. Moreover, we map $appreciations_c$ to the amount of feedback obtained by these contributions:
        \begin{equation}
            q_v = \frac{ 
        	    \sum\limits_{c_1 \in Revs_v \cup Tips_v} useful_{c_1}+fun_{c_1}+cool_{c_1}+like_{c_1}
            }{\max\limits_{a\in {\cal U}}\sum\limits_{c_2 \in Revs_a\cup Tips_a} useful_{c_2}+fun_{c_2}+cool_{c_2}+like_{c_2}}
        \end{equation} 
    \end{enumerate}
\end{itemize}

\subsection{Instantiation of MTM in a different application domain}
Le'ts consider, as a further example of instantiation of MTM, the \cite{LibraryThing} social network that publishes information about books. LibraryThings enables its members to create their own virtual libraries and to tag and review books. Users can establish friend relations to watch and take inspiration from the libraries created by other people; moreover, they can visualize the reviews published in the social network and they can express positive-only feedback about the helpfulness of each review. Users are not enabled to endorse other users' profiles.
LibraryThing discloses the social relations among users and the number of helpfulness votes received by each review.
Trust indicators can be mapped to MTM trust classes as follows:
\begin{itemize}
\item 
{\em Quality of individual contributions on an item } ($fContr_{vi}$).
For this class of trust evidence we apply Equation \ref{eq:fPositiveFeedback} of page \pageref{eq:fPositiveFeedback}, which is suitable for positive-only feedback. For each review $r$ published in the social network, we thus map $appreciations_{r}$ to the number of helpful votes that $r$ has received.
\item 
{\em Multi-dimensional global reputation} ($mgr_v$). We compute $mgr_v$ by fusing in Equation \ref{eq:multi-dimensional-rep} the $imp_v$ and $q_v$ indicators respectively describing {\em v}'s importance in the social network and her/his quality as a contributor:
    \begin{equation}
    mgr_v = \frac{C_1imp_v+C_6q_v}
    {C_1+C_6}
    \end{equation}
We can compute $imp_v$ as $v$'s PageRank score by transforming bidirectional friend relations to pairs of unidirectional social links. Moreover, $q_v$ can be defined as the ratio between the total number of helpful votes received by {\em v}'s reviews and the maximum number of helpful votes received by the other members of the social network.
\end{itemize}

\section{Validation of LOCABAL+}
\label{sec:validation}

\subsection{Evaluation metrics}
\label{sec:metrics}
As mentioned in Section \ref{sec:questions}, we evaluate recommendation algorithms on the basis of {\em accuracy} and {\em error minimization} (i.e., the ability to provide correct results), {\em ranking capability} (i.e., the ability to correctly sort items depending on their ground truth relevance to the user) and {\em user coverage} (i.e., the percentage of users for whom the recommender is able to find items that are likely to be relevant). 
This is in line with the recent trends in the evaluation of recommender systems, which do not exclusively focus on accuracy to provide a broader view on performance; e.g., see \citep{Jannach-etal:16}.
Before describing the evaluation metrics in detail we introduce the notation we use:
\begin{itemize}
\item 
$U$ is the set of users and $I$ the set of items; $R$ is the set of ground truth ratings and $\hat{R}$ the set of estimated ones.
\item 
$r_{ui}$ is the rating score that $u \in U$ has given to $i \in I$ and $\hat{r}_{ui}$ is the rating score estimated by the recommender system.
\item 
$Relevant_u$ is the set of items that $u$ has positively rated; in a [1, 5] Likert scale we define $Relevant_u = \{i \in I ~|~ r_{ui}>3\}$.
\item 
$Recommended_u$ is the set of items that the system suggests to $u$: $Recommended_u = \{i \in I ~|~ \hat{r}_{ui}>3\}$.
\end{itemize}
We evaluate recommendation accuracy and error minimization by means of the following metrics: 
\begin{itemize} 
\item 
Precision: P@k = ${\frac  {1}{|U|}} \sum\limits_{u \in U} \text{P}_u$@k, where P$_u$@k = $\frac{ |Recommended_u \cap Relevant_u| }{ |Recommended_u|}$
\item 
Recall: R@k = ${\frac  {1}{|U|}} \sum\limits_{u \in U} \text{R}_u$@k, where R$_u$@k = $\frac{|Recommended_u \cap Relevant_u| }{ |Relevant_u| }$
\item 
Accuracy: F1@k $= 2*\frac{  \text{P@k * R@k}  }{  \text{P@k + R@k} }$ 
\item 
Root Mean Squared Error: RMSE@k = $\sqrt{\frac{1}{|\hat{R}@\text{k}|} \sum\limits_{\hat{r}_{ui} \in \hat{R@\text{k}}}(r_{ui} - \hat{r}_{ui})^2}$
\item 
Mean Absolute Error: MAE@k = $\frac{1}{|\hat{R}@\text{k}|} \sum\limits_{\hat{r}_{ui} \in \hat{R@\text{k}}}|r_{ui} - \hat{r}_{ui}|$
\end{itemize}
As far as ranking capability is concerned we use the following metrics:
\begin{itemize} 
\item 
Mean Reciprocal Rank, which measures the placement of the first relevant items in recommendation lists:

MRR@k = ${\frac  {1}{|U|}}\sum\limits_{u \in U} {\frac  {1}{rank_{u}}}$, where ${rank}_{u}$ is the position of the first relevant item in the list generated for user $u$.

\item 
Mean Average Precision, which measures the average correct positioning of items in the recommendation lists:

MAP@k = ${\frac  {1}{|U|}} \sum\limits_{u \in U} {\frac  {1}{|Relevant_u|}} \sum _{{x=1}}^{\text{k}}\text{P}_u\text{@x} * Rel_u(x)$

where $Rel_u(x) = 1$ if the item in position $x$ of the list for $u$ is relevant to her/him, 0 otherwise.
\end{itemize}
Finally, we measure User Coverage (shortened to UCov in the tables showing the evaluation results) as the percentage of users of the dataset for whom the algorithm finds at least one item $i \in I$ such that $\hat{r}_{ui}>3$, i.e., an item that the system evaluates as relevant to the user. 

\begin{table*}[t]
\centering 
\caption{Configurations of LOCABAL+ used in the experiments.}
\resizebox{\columnwidth}{!}{%
{\def\arraystretch{1.5}
\begin{tabular}{lccc}
\hline
Configuration & Social relations & Feedback on user profiles & Feedback on user contributions  \\ 
\hline
LOCABAL+ & Yes & Yes & Yes   \\
LOC+noF & Yes & Yes & -   \\
LOC+noE & Yes & - & Yes   \\
LOC+noS & - & Yes & Yes \\
\hline
\end{tabular}
}}
\label{tab:configurations}
\end{table*}

\subsection{Methodology applied in the experiments}
We consider various configurations of MTM to evaluate the performance of LOCABAL+ when using all the facets of trust available in the YELP-Hotel/Yelp-Food datasets, or a subset of them. We are interested in understanding whether the algorithm can provide good recommendation results when we omit different sources of evidence in order to assess its applicability to social networks that disclose different types of information about users. Specifically, we consider the following cases, summarized in Table \ref{tab:configurations}:
\begin{itemize} 
\item 
LOCABAL+. This is the algorithm applied to the complete information available in the dataset (social relations, feedback about users and feedback about contributions). It computes multi-dimensional global reputation with $C_1 = \dots = C_6 = 1$ in Equation \ref{eq:mgr_Yelp}, and multi-faceted trust for item $i$ with $C = 1$ in Equation \ref{eq:mt}.

\item 
LOC+noF. This configuration ignores the feedback about reviews and tips; thus, it only relies on multi-dimensional global reputation, which is computed by taking social links and global feedback on user profiles into account. In detail, LOC+noF is obtained by switching off the quality of the user as a contributor ($q_v$) in the computation of multi-dimensional global reputation ($C_6 = 0$  in Equation \ref{eq:mgr_Yelp}) and the feedback received by the specific contribution ($fContr_{vi}$) in the computation of multi-faceted trust for item $i$ ($C = 0$ in Equation \ref{eq:mt}).
LOC+noF is useful to understand whether, by only using social information and anonymous feedback on user profiles, the recommender system is able to generate useful suggestions.
\item 
LOC+noE. This configuration ignores the global feedback on user profiles, i.e., the trust indicators of class U in Section 
\ref{sec:MGR} (user profile endorsements and public recognition, visibility) in the computation of multi-dimensional global reputation (i.e., $C_2 = C_3 = C_4 = C_5 = 0$ in Equation \ref{eq:mgr_Yelp}).
LOC+noE is particularly interesting because not all of the social networks manage profile endorsements; e.g., we mentioned in Section \ref{sec:datasets} that LibraryThing does not support this type of feedback. Therefore we are interested in understanding whether the recommender system can achieve good performance by only relying on social links and feedback on user contributions.
\item 
LOC+noS. This configuration ignores social relations; it is obtained by switching off the importance of users ($imp_v$) in the computation of multi-dimensional global reputation ($C_1 = 0$ in Equation \ref{eq:mgr_Yelp}) and the social regularization component of LOCABAL+ ($\alpha=0$ in Equation \ref{eq:LOCABAL+}).
LOC+noS helps understand whether, thanks to the exploitation of public, anonymous feedback about users and user contributions, LOCABAL+ can generate good recommendations in the application domains where the information about social links is unavailable. As previously discussed, this is an important aspect for the applicability of trust-based recommender systems, given the growing sensibility of users towards privacy protection.

\end{itemize}
In the experiments we use the Surprise \citep{Surprise} implementation of U2UCF and SVD++ and the RecQ implementation of LOCABAL \citep{RecQ}. 
LOCABAL+ and U2USocial are developed by extending the implementations of LOCABAL and U2UCF respectively.
All the algorithms are integrated in Surprise to uniformly evaluate their performance. 

On each dataset we organize the evaluation as follows:
we first validate the algorithms on 90\% of the dataset by running Grid Search to find the best configuration of parameters with respect to MAP, using 5 cross-fold validation. All the executions are performed having set 50 latent factors.
Then we additionally test the best configuration obtained from Grid Search on the remaining 10\% of the dataset to measure the performance of the algorithms on new data in order to check their impact in a dynamic environment where new ratings are continuously provided.

\begin{table*}[t]
\centering 
\caption{Performance@10 on Yelp-Hotel dataset (the best results are in boldface). The ``$\diamond$'' symbol means that results are significant at $p<0.05$ with respect to all the baselines except for SocialMF; the ``$\dagger$'' means that results are significant at $p<0.01$ with respect to all the baselines except for U2UCF.}
\resizebox{\columnwidth}{!}{%
{\def\arraystretch{1.3}
\begin{tabular}{lcccccccccccc}
\hline
 & $\alpha$ & $\beta$ & P & R & F1 & MAP & RMSE & MAE & MRR & UCov \\ 
\hline
Significance  & - & - & 0.01 & - & 0.01 & $\dagger$ & 0.02 & 0.02 & - & $\diamond$ \\
\hline
LOCABAL+      & 0.9 & 0   & 0.7919 & 0.7389 & {\bf 0.7645} & {\bf 0.5303} & 0.8922 & 0.671  & {\bf 0.6125} & 0.7543  \\
LOC+noF   & 0.1 & 0.3 & 0.7923 & 0.7381 & 0.7642 & 0.5288 & 0.8927 & 0.6708 & 0.6087 & 0.7523 \\
LOC+noE    & 0.3               & 0.3              & 0.7923     & 0.7377     & 0.764       & 0.5285       & 0.8938        & 0.6716       & 0.6086       & 0.7517       \\
LOC+noS   & - & 0.1 & {\bf 0.7931} & 0.7368 & 0.7639 & 0.5274 & {\bf 0.8916} & {\bf 0.6702} & 0.6082 & 0.7513   \\
U2UCF    & -   & -   & 0.76   & {\bf 0.7399} & 0.7498 & 0.5215 & 0.9582 & 0.7264 & 0.5982 & 0.6127  \\
SocialMF & -   & -   & 0.7757 & 0.7261 & 0.7501 & 0.5116 & 0.9238 & 0.6954 & 0.6055 & 0.7655  \\
LOCABAL      & 0.1 & -   & 0.7732 & 0.7259 & 0.7488 & 0.5112 & 0.9281 & 0.6994 & 0.6078 & 0.7698 \\
SVD++    & -   & -   & 0.7595 & 0.717  & 0.7376 & 0.4994 & 0.976  & 0.7383 & 0.5993 & {\bf 0.7755}  \\
U2USocial      & -   & -   & 0.7503 & 0.7233 & 0.7366 & 0.4798 & 1.0085 & 0.773 & 0.5336 & 0.2589   \\
\hline
\end{tabular}
}
}
\label{tab:performance-Hotel}
\end{table*}

\subsection{Evaluation results}
\label{sec:evaluationResults}
\subsubsection{Yelp-Hotel}
\label{sec:Yelp-Hotel-results}
Table \ref{tab:performance-Hotel} compares the performance achieved by each configuration of LOCABAL+ to that of the baselines (U2UCF, SocialMF, LOCABAL, SVD++ and U2USocial) on Yelp-Hotel. In this table, as well as in the following ones, the best values are in boldface. The significance level of results, reported in the second row, is obtained by separately comparing each configuration with all the baselines. The rows describing performance are sorted by MAP, from the best one to the worst one, in order to highlight the ranking capabilities of the algorithms.

As shown in the central portion of the table, in this dataset the LOCABAL+ configurations outperform the baselines in all measures except for (i) Recall that is dominated by U2UCF, and (ii) User Coverage, where SVD++ is the best algorithm, followed by LOCABAL and SocialMF. The loss in user coverage is however compensated by higher accuracy and ranking capability as LOCABAL+ is the best algorithm in terms of F1, MAP and MRR. Noticeably, LOC+noS is the most precise algorithm, excelling in P, RMSE and MAE. In the following we analyze the LOCABAL+ configurations.

LOCABAL+ obtains its best performance with $\alpha=0.9$ and $\beta=0$. 
The value of $\alpha$ shows that the algorithm strongly relies on users' multi-dimensional reputation to steer social regularization: instead of minimizing the impact of the local social context of users ({\bf S$_{xz}$} in Equation \ref{eq:LOCABAL+}), as done in LOCABAL (where $\alpha=0.1$ uniformly flattens the impact of the local social context across users), LOCABAL+ tunes social regularization on the basis of the preferences of the most similar and reputable friends. 
Differently, $\beta=0$ means that the multi-faceted trust $mtf_{xy}$ that tunes the impact of ratings in Matrix Factorization is computed by ignoring users' reputation; therefore, for this purpose, the algorithm only relies on the feedback ($fContr_{vi}$) received by the reviews and tips associated with the ratings. This is different from LOCABAL, which tunes the impact of ratings on the basis of users' PageRank score.   
In summary, in Yelp-Hotel, LOCABAL+ steers social regularization by multi-dimensional reputation and weights the impact of ratings on the basis of the publicly recognized value of user contributions, which emerges as a good source of information to identify reliable ratings.

It should however be noticed that, in LOCABAL+, multi-dimensional global reputation is computed by taking multiple types of trust evidence into account, i.e., PageRank score, user profile endorsements and quality of the user as a contributor that, in turn, derives from the feedback on contributions. Therefore, it is difficult to say which type of evidence brings the most useful information. In order to clarify the situation we analyze the other configurations of LOCABAL+.

LOC+noF ignores the feedback on user contributions and is optimized with $\alpha = 0.1$ and $\beta=0.3$. It has lower performance than LOCABAL+ but it outperforms all the baselines in Precision, F1, MAP, RMSE, MAE and MRR. The value of $\alpha$ dramatically weakens the role of social regularization in the Matrix Factorization process with respect to LOCABAL+ (it is flattened to 10\% as in LOCABAL, but it is much weaker due to the presence of the $mgr_z$ term within the nested summation of Equation \ref{eq:LOCABAL+}). Moreover, $\beta=0.3$ means that the values of multi-faceted trust computed by the algorithm are reduced to 30\% of the multi-dimensional reputation. However, given the weak role of social regularization, reputation is central to learning the user preference and item characteristic vectors. These findings are coherent with the hypothesis that, in this dataset, the feedback on user contributions is a very useful type of information to learn user preferences but show that, even by only employing social links and the feedback on user profiles, the algorithm can achieve satisfactory results.

LOC+noE ignores the feedback on user profiles. It obtains its best performance with $\alpha = \beta = 0.3$: with respect to LOCABAL+, the algorithm weights social regularization much less but it partially takes multi-dimensional reputation into account (30\%) in the computation of multi-faceted trust. With respect to LOC+noF, LOC+noE increases a little bit the role of social regularization in the Matrix Factorization process. The algorithm outperforms the baselines and is generally worse than LOCABAL+. Moreover, it performs slighly worse than LOC+noF: it has the same precision and very similar F1, MAP and MRR but it has lower recall, RMSE, MAE and User Coverage. We explain these findings with the fact that, as LOC+noE ignores the trust feedback received by user profiles, it generally misses useful information for preference prediction. In the dataset, user endorsements have high median values (e.g., the median number of compliments received by individual users is 110 and the median number of fans is 41). Therefore, when the algorithm ignores them, it has fewer chances to recognize highly reputable users. The value of $\beta$ also shows that the feedback on user contributions determines the value of multi-faceted trust by 70\%. Once more, it looks like the feedback on user contributions has an important role in defining trust; however, the feedback on user profiles is useful as well.

LOC+noS ignores social relations among users: it only employs anonymous trust statements and anonymous social information (number of fans) to learn user preferences. This means that it is not possible to compute the importance of users in the social network ($imp_v$) and that social regularization does not make sense. We obtain LOC+noS by forcing parameter $C_1 = 0$ in Equation \ref{eq:mgr_Yelp} and $\alpha=0$ in Equation \ref{eq:LOCABAL+}.  This algorithm outperforms the baselines in all measures except for User Coverage and Recall; moreover, it has the best RMSE and MAE of all the algorithms and configurations of LOCABAL+. This supports the hypothesis that, in this dataset, anonymous trust feedback is a precious source of information to be used in a recommender system, and that correct ratings can be predicted without using personal data about social relations.

\begin{figure*}[t]
\centering
\captionsetup[subfigure]{position=b}
 \begin{subfigure}[t]{0.49\textwidth}
 \centering
 \includegraphics[width=1\linewidth]{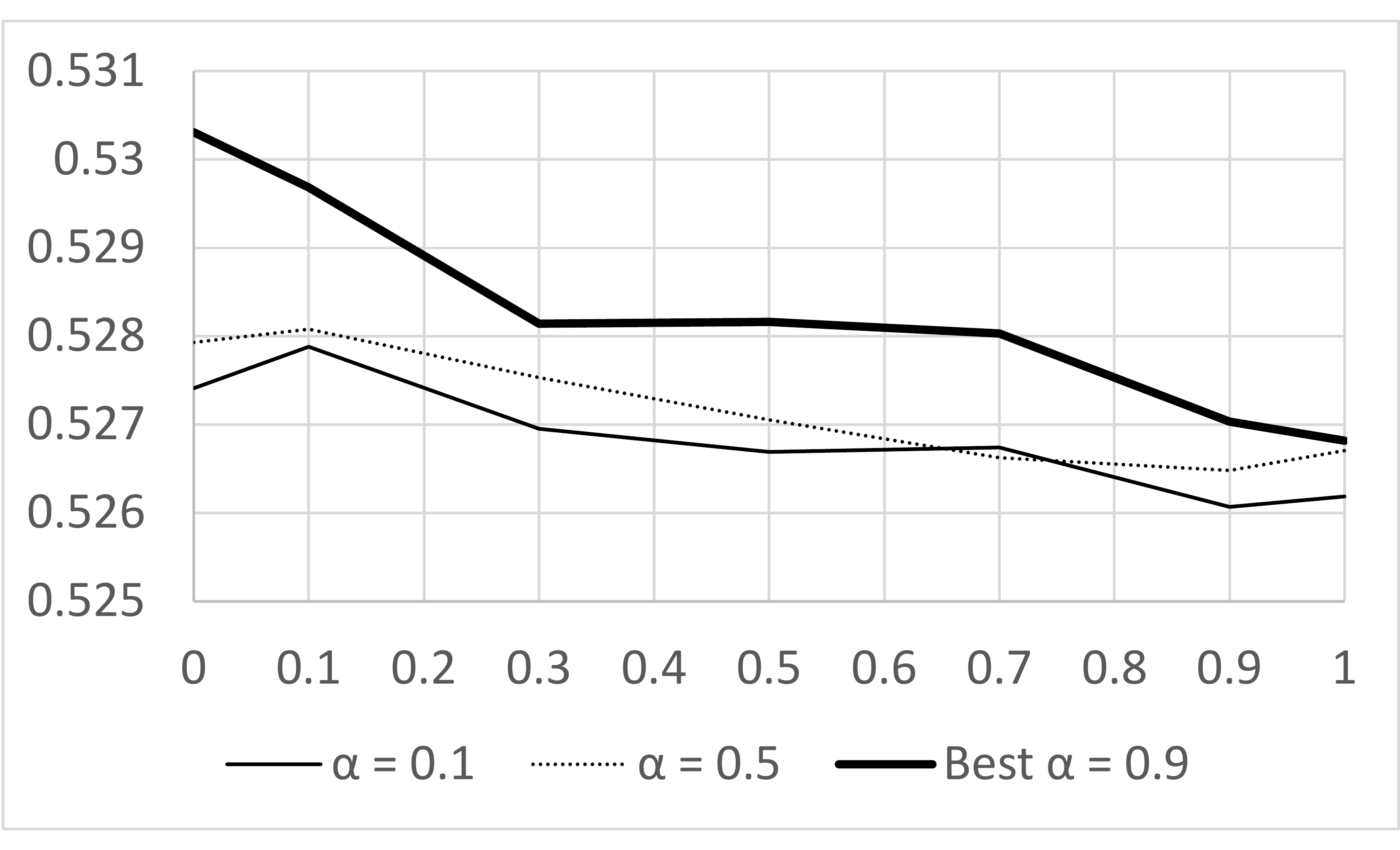}
 \caption{MAP variation with respect to $\beta$}\label{fig:graficiHotelBeta}
\end{subfigure}
\hfill
\begin{subfigure}[t]{0.49\textwidth}
 \centering
 \includegraphics[width=1\linewidth]{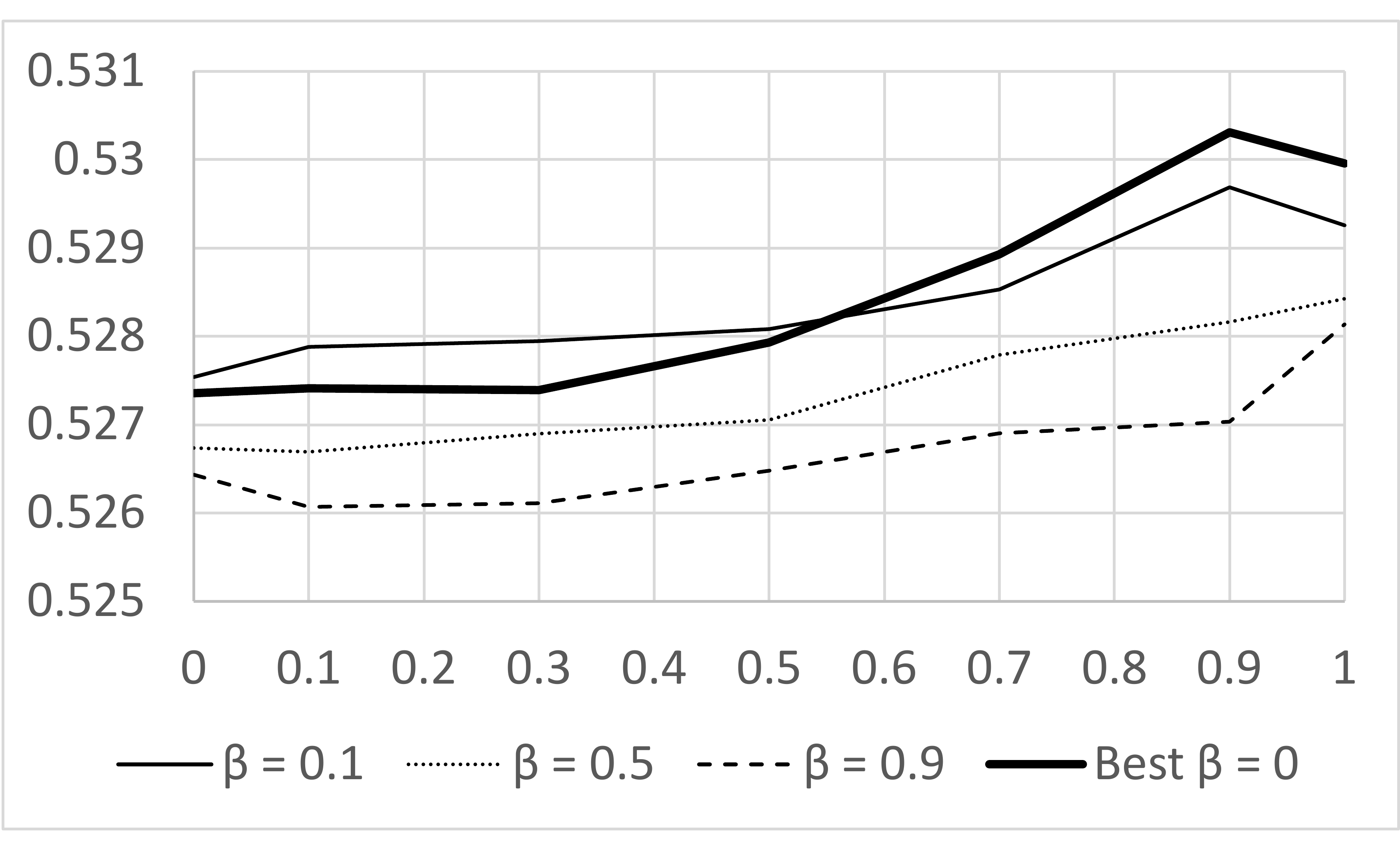}
 \caption{MAP variation with respect to $\alpha$}\label{fig:graficiHotelAlpha}
\end{subfigure}

\caption{MAP variation on Yelp-Hotel. The Y axis represents MAP; the X axis represents $\beta$ in Figure \ref{fig:graficiHotelBeta} and $\alpha$ in the Figure \ref{fig:graficiHotelAlpha}.}\label{fig:graficiHotel}
\end{figure*}

Figure \ref{fig:graficiHotel} shows the variation of MAP for all the configurations of LOCABAL+, depending on $\alpha$ and $\beta$.
By setting $\alpha$ to a constant value (Figure \ref{fig:graficiHotelAlpha}), MAP decreases when $\beta$ grows. This means that, having blocked the influence of the social component of LOCABAL+, the best results are achieved when the feedback on contributions makes ratings more influent on the Matrix Factorization process. Moreover, by setting $\beta$ to a constant value (Figure \ref{fig:graficiHotelBeta}), MAP improves when $\alpha$ increases and the best results are achieved with $\alpha$=0.9, i.e., when the social component, in combination with the other facets of trust, strongly influences preference estimation.

\begin{table*}[t]
\centering 
\caption{Performance@10 on the new data of the Yelp-Hotel dataset.}
\resizebox{\columnwidth}{!}{%
{\def\arraystretch{1.3}
\begin{tabular}{lcccccccccccc}
\hline
 & $\alpha$ & $\beta$ & P & R & F1 & MAP & RMSE & MAE & MRR & UCov \\ 
\hline
LOCABAL+ & 0.9 & 0  & {\bf 0.8253}     & {\bf 0.7703}     & {\bf 0.7968}      & {\bf 0.4818}       & {\bf 0.8801}        & {\bf 0.6601}       & 0.5341       & 0.6748        \\
SocialMF  & - & - & 0.8044     & 0.7687     & 0.7862      & 0.4808       & 0.918         & 0.701        & {\bf 0.5399} & 0.7055        \\
LOCABAL & 0.1 & -  & 0.8041     & 0.761      & 0.782       & 0.4736       & 0.9369        & 0.7158       & 0.5378       & 0.7014        \\
U2UCF & - & -    & 0.7906     & 0.7684     & 0.7794      & 0.4723       & 0.9627        & 0.7251       & 0.5239       & 0.5767        \\
SVD++ & - & -    & 0.7867     & 0.7607     & 0.7735      & 0.4709       & 0.9832        & 0.7457       & 0.5324       & {\bf 0.7076}        \\
U2USocial & - & - & 0.7759     & 0.7561     & 0.7658      & 0.4518       & 1.0597        & 0.7975       & 0.4952       & 0.2802   \\
\hline
\end{tabular}
}
}
\label{tab:Yelp-Hotel-extra}
\end{table*}

Table \ref{tab:Yelp-Hotel-extra} summarizes the evaluation results on new data (10\% of Yelp-Hotel). It can be seen that the results are fairly consistent with those of Table \ref{tab:performance-Hotel}: LOCABAL+ outperforms the baselines in all performance measures except for User Coverage, as previously, dominated by SVD++ and SocialMF. The main difference in this case is that SocialMF also dominates MRR. We can conclude that LOCABAL+ can be used in dynamic application domains without frequently optimizing the model.

\begin{table*}[t]
\centering 
\caption{Performance@10 on Yelp-Food dataset.}
\resizebox{\columnwidth}{!}{%
{\def\arraystretch{1.3}
\begin{tabular}{lcccccccccccc}
\hline
 & $\alpha$ & $\beta$ & P & R & F1 & MAP & RMSE & MAE & MRR & UCov \\
\hline
Significance  & & & 0.01 & 0.01 & 0.01 & 0.01 & 0.01 & 0.01 & 0.04 & 0.01 \\
\hline

LOCABAL+   & 0.7               & 0.3              & 0.7769     & {\bf 0.7528}     & 0.7647      & {\bf 0.5993}       & 0.9791        & 0.7406       & {\bf 0.7166}       & 0.8499        \\
LOC+noE    & 0.9               & 0.3              & 0.7769     & 0.7527     & 0.7646      & 0.5993       & 0.9793        & 0.7407       & 0.7165       & 0.8497  \\
LOC+noS    & -                 & 0.3              & 0.777      & 0.7527     & 0.7647      & 0.5992       & 0.9789        & 0.7403       & 0.7164       & 0.8497        \\
LOC+noF    & 0.5               & 0.7              & {\bf 0.7783}     & 0.752      & {\bf 0.7649}      & 0.5986       & {\bf 0.9768}        & {\bf 0.7389}       & 0.7163       & 0.8485         \\ 
SocialMF   & -                 & -                & 0.76       & 0.7312     & 0.7453      & 0.5741       & 1.0358        & 0.7828       & 0.7092       & 0.8642        \\
LOCABAL    & 0.1               & -                & 0.7586     & 0.7294     & 0.7437      & 0.5724       & 1.041         & 0.7866       & 0.7093       & 0.8656        \\
U2UCF      & -                 & -                & 0.7488     & 0.7337     & 0.7412      & 0.5642       & 1.0796        & 0.8072       & 0.6795       & 0.7025        \\
SVD++      & -                 & -                & 0.7483     & 0.7116     & 0.7295      & 0.5539       & 1.0761        & 0.812        & 0.7017       & {\bf 0.8698}        \\
U2USocial & -                 & -                & 0.7729     & 0.7357     & 0.7539      & 0.5474       & 1.0741        & 0.8178       & 0.6061       & 0.2295 \\       
\hline
\end{tabular}
}
}
\label{tab:performance-Food}
\end{table*}

\subsubsection{Yelp-Food}
Table \ref{tab:performance-Food} shows the evaluation results on the Yelp-Food dataset. The LOCABAL+ configurations outperform the baselines with statistically significant results in all measures except for User Coverage. Moreover, SVD++ has the highest coverage and LOCABAL is the second best, followed by SocialMF, but all of them are less accurate and have lower ranking capability than the LOCABAL+ configurations. In this case, LOCABAL+ and LOC+noF are the best performing algorithms but, as LOCABAL+ has the best Recall, MAP and MRR (and LOC+noF the worst ones among LOCABAL+ configurations), we consider LOCABAL+ as the preferable one.

In this dataset, all the LOCABAL+ configurations take multi-dimensional global reputation into account in the computation of multi-faceted trust. Specifically, $\beta=0.3$ in LOCABAL+, LOC+noE and LOC+noS; moreover, $\beta=0.7$ in LOC+noF that overlooks the feedback on user contributions. Moreover, social regularization has a medium to high role in the Matrix Factorization process, with the strongest influence in LOCABAL+ ($\alpha=0.7$) and LOC+noE ($\alpha=0.9$). In summary, the configurations exploit both multi-faceted trust and social regularization to obtain their best performance but the exclusion of feedback on user contributions raises the importance of the other facets of trust.
This can be explained by the fact that users and user contributions receive a low amount of feedback for the evaluation of trustworthiness; e.g., the median number of appreciations is 1 for reviews and 0 for tips and the median number of endorsements to user profiles, including Elite years, compliments and fans, is 6. Thus, the algorithms rely on the joint contribution of all the sources of evidence about trust. 
Another relevant observation is that the performance of LOC+noE and LOC+noS is similar to that of LOCABAL+. This can be explained by assuming that, in this dataset, the feedback on user profiles and social relations play complementary roles and can replace each other without a major loss of performance.

\begin{figure*}[t]
\centering
\captionsetup[subfigure]{position=b}
 \begin{subfigure}[t]{0.49\textwidth}
 \centering
 \includegraphics[width=1\linewidth]{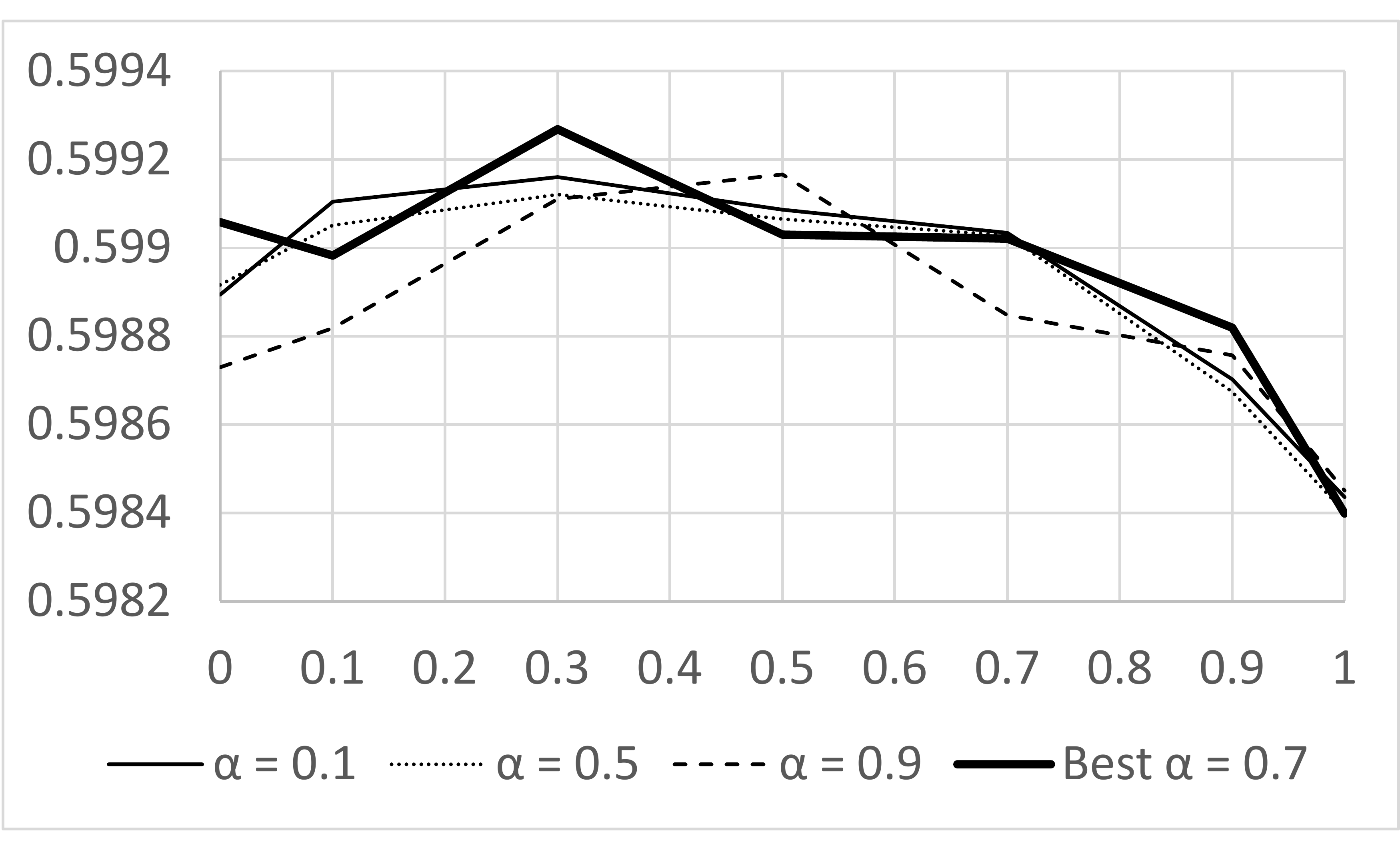}
 \caption{MAP variation with respect to $\beta$. }\label{fig:graficiFoodBeta}
\end{subfigure}
\hfill
\begin{subfigure}[t]{0.49\textwidth}
 \centering
 \includegraphics[width=1\linewidth]{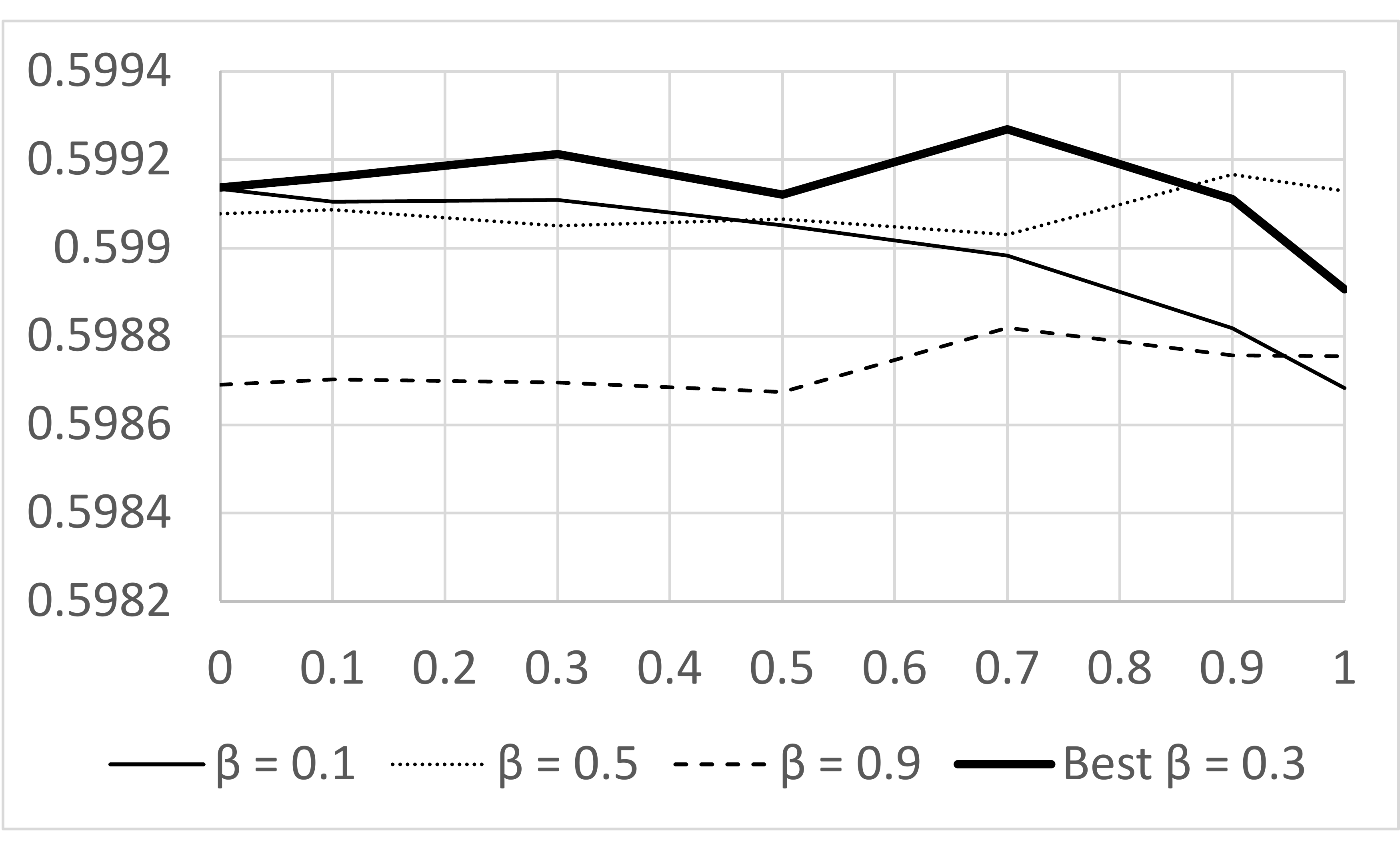}
 \caption{MAP variation with respect to $\alpha$.}\label{fig:graficiFoodAlpha}
\end{subfigure}
\caption{MAP variation on Yelp-Food. The Y axis represents MAP; the X axis represents $\beta$ in Figure \ref{fig:graficiFoodBeta} and $\alpha$ in Figure \ref{fig:graficiFoodAlpha}.}
\label{fig:graficiFood}
\end{figure*}

Figure \ref{fig:graficiFood} shows the variation of MAP for all the LOCABAL+ configurations depending on $\alpha$ and $\beta$. 
It can be noticed that by setting $\alpha$ to a constant value (Figure \ref{fig:graficiFoodAlpha}), MAP first slightly increases but, when $\beta>0.3$, it quickly decreases. This means that, regardless of the influence of the social component of LOCABAL+, the algorithm benefits from a moderate support by multi-dimensional global reputation.
Notice also that, by setting $\beta$ to a constant value (Figure \ref{fig:graficiFoodBeta}), results are not particularly affected by the value of $\alpha$; therefore, social regularization has a generally constant contribution to recommendation performance.

\begin{table*}[t]
\centering 
\caption{Performance@10 on the new data of the Yelp-Food dataset.}
\resizebox{\columnwidth}{!}{%
{\def\arraystretch{1.3}
\begin{tabular}{lcccccccccccc}
\hline
 & $\alpha$ & $\beta$ & P & R & F1 & MAP & RMSE & MAE & MRR & UCov \\
\hline
LOCABAL+ & 0.7 & 0.3  & {\bf 0.7935}     & {\bf 0.787} & {\bf 0.7903}      & {\bf 0.5835}       & {\bf 0.9701}        & {\bf 0.7376}       & {\bf 0.6584}   & {\bf  0.8053}        \\
U2UCF & - & -    & 0.769      & 0.7699     & 0.7694      & 0.5606       & 1.0423        & 0.7807       & 0.6378       & 0.7193        \\
U2USocial & - & - & 0.7798     & 0.7708     & 0.7753      & 0.5472       & 1.0568        & 0.8055       & 0.5913       & 0.2595        \\
SocialMF & - & - & 0.7818     & 0.7489     & 0.765       & 0.5461       & 1.0403        & 0.7829       & 0.6387       & 0.8011        \\
LOCABAL & 0.1 & -  & 0.7814     & 0.7478     & 0.7642      & 0.5451       & 1.0449        & 0.7864       & 0.6384       & 0.8025        \\
SVD++ & - & - & 0.7742     & 0.7389     & 0.7561      & 0.5339       & 1.0712        & 0.8056       & 0.6314       & 0.8033   \\
\hline

\end{tabular}
}
}
\label{tab:Yelp-Food-extra}
\end{table*}

Table \ref{tab:Yelp-Food-extra} shows the evaluation of algorithms on new data. It can be seen that LOCABAL+ outperforms the baselines in all performance metrics, including MRR. Thus, we can conclude that also in Yelp-Food LOCABAL+ can be employed in dynamic environments without requiring a frequent optimization of parameters.

\section{Discussion and future work}
 \label{sec:discussion}
The experimental results show that, in the Yelp-Hotel and Yelp-Food datasets, LOCABAL+ outperforms all the baseline recommender systems in accuracy, error minimization and ranking capability with a minor loss of User Coverage with respect to SVD++, LOCABAL and SocialMF. Thus, we can say that LOCABAL+ generates better suggestions to marginally fewer people, with a clear positive gain in performance.

Noticeably, LOCABAL+ outperforms the baselines even though MTM is configured to ignore different types of evidence about trust; this finding shows the flexibility of our approach towards the lack of user information. This is generally important because some classes of trust information defined in MTM might not be available in specific recommendation domains. Moreover, it is a key achievement in relation to the management of personal data because, different from the other trust-based recommender systems, LOCABAL+ can work by ignoring data about social links, and by only relying on public anonymous information. This aspect is more and more important given the increasing sensibility of users towards disclosing personal data.

The evaluation results on new data confirm that, on both datasets, LOCABAL+ achieves the best performance in all the measures except for User Coverage (actually, in Yelp-Food LOCABAL+ outperforms the other algorithms in User Coverage as well, but this is not true in Yelp-Hotel). Thus, we conclude that LOCABAL+ can be applied, without losing recommendation capability, to dynamic environments in which it can not be frequently optimized.

The evaluation results are useful to answer our research questions:
\begin{itemize}
    \item[RQ1:]
    {\em Can multi-faceted trust be used to improve the performance of a trust-based recommender system with respect to only relying on social links and rating similarity among users?}
    Our experiments enable us to positively answer RQ1 because they provide consistent results on two datasets having different characteristics (e.g., size, distributions of global feedback on users and contributions, etc.); see Section \ref{sec:datasets}. Specifically, the superior results of LOCABAL+ with respect to the baselines and, in particular, to LOCABAL, show that by exploiting multiple facets of trust (including global anonymous feedback on users and user contributions), the performance of the recommender system significantly improves.  It is worth noticing that, thanks to MTM, trust-based methods can be successfully applied without using personal information about social links. This fact represents a key aspect of our approach compared to existing work on trust-based recommender systems. 
    \item[RQ2:]
    {\em What is the impact of the multi-dimensional reputation of users, of the quality of their contributions, and of social links, on collaborative recommendation performance?}
    Our experiments show that LOCABAL+ can work without using data about social relations, with a minor loss of performance, if it can use other types of trust feedback.
    Regarding the other facets of trust, the experiments show that their relative importance depends on the amount and quality of the feedback about users and user contributions available to the recommender system. The two datasets we selected are interesting because the diverse distributions of feedback they provide determine slightly different behavior of the LOCABAL+ configurations.
    In Yelp-Hotel, which provides a large amount of feedback about user contributions (median = 53 against 27 of Yelp-Food), the algorithm obtains the best performance results by privileging this type of information over multi-dimensional global reputation; moreover, performance results are clearly affected by the omission of the feedback on user contributions. Differently, in the Yelp-Food dataset, which stores scarser feedback about contributions, the algorithm obtains the best results by balancing the influence of all types of global feedback, including the endorsements to user profiles, and performance increases if feedback on user contributions is ignored.
\end{itemize}
We thus conclude that global feedback about user trust is a useful type of information in Top-N recommendation. However, in order to decide which types of evidence should be used, it is important to analyze the characteristics of the data to which the recommender system is applied and the features of the various types of feedback that can be used.

Before closing this section it is worth noting that our current work focuses on numerical aspects of the evidence about trust, by measuring the amount of feedback that users and user contributions receive. However, the content of the global feedback provided by users is itself a further source of information about trust that can be analyzed to acquire information about users' behavior. In this perspective, we plan to extend our model to the analysis of the content of reviews (and/or microblogs), which has been largely studied to evaluate their quality \citep{Huang-etal:15,Mudambi-Schuff:10,Chua-Banerjee:15,Qazi-etal:16,Korfiatis-etal:12,Krishnamoorthy:15,Kim-etal:06}, to steer personalized recommendation \citep{Raghavan-etal:12,Alhamadi-Zeng:15,Shen-etal:2019,Rubio-etal:19} and to guide the explanation of recommendations \citep{O'Mahony-Smyth:18,Musto-etal:2019}.

\section{Conclusions}
\label{conclusions}
This paper has described the Multi-faceted Trust Model (MTM) and the LOCABAL+ recommender system, which combine social links with global anonymous feedback about users and user contributions to enhance collaborative Top-N recommendation. LOCABAL+ extends the LOCABAL recommender system with multi-faceted trust and with an enhanced regularization of social relations based on both rating similarity and users' multi-dimensional global reputation. 

LOCABAL+ has various advantages with respect to the state-of-the art trust-based recommender systems. In particular, being based on the MTM compositional model, it can be configured to work with different types of evidence about trust, such as anonymous public feedback on user profiles and user contributions, as well as information about social relations. Interestingly, LOCABAL+ can work by ignoring the information about social links, which has been recently found to be problematic in the user acceptance of trust-based recommender systems because it is considered as personal information. 
Another advantage of LOCABAL+ is that the extension to social regularization makes it possible to select users to steer Matrix Factorization in a more selective way with respect to only considering rating similarity.

Experiments carried out on two public datasets of item reviews show that, with a minor loss of user coverage, LOCABAL+ outperforms state-of-the art trust-based recommender systems and Collaborative Filtering in accuracy, error minimization and ranking capability both when it uses complete information from the datasets and when it ignores social relations (or other types of feedback on users and contributions). It thus represents a flexible approach to trust-based recommendation, suitable to comply with specific data management requirements.

\section{Acknowledgments}
This work was supported by the University of Torino which funded the conduct of the research and preparation of the article.

\bibliographystyle{elsarticle-harv}

\newpage
\section{Author contributions}
Liliana Ardissono has the following contributions in the present paper:
Conceptualization; 
Funding acquisition; Methodology; Resources;
Supervision; Roles/Writing - original draft.

Noemi Mauro has the following contributions in the present paper:
Conceptualization; Data curation;
Investigation; Methodology; 
Software;  Validation; Roles/Writing - original draft.

\end{document}